\documentclass[usenatbib]{aa}  
\usepackage{graphicx}
\usepackage{float}

\usepackage{txfonts}
\usepackage{xcolor}

\newcommand{\bz}{$\langle B_z \rangle$}
\newcommand{\nz}{$\langle N_z \rangle$}
\newcommand{\vsini}{$v \sin i$}
\newcommand{\kms}{km\,s$^{-1}$}

\newcommand{\mdot}{$\dot{M}$}

\newcommand{\msun}{M$_\odot$}

\newcommand{\teff}{$T_{\rm eff}$}
\newcommand{\ra}{$R_{\rm A}$}
\newcommand{\rk}{$R_{\rm K}$}

\begin{document} 

   \title{Discovery of the binary nature of the magnetospheric B-type star $\rho$\,Oph\,A}
   \titlerunning{Spectropolarimetry of $\rho$\,Oph\,A}

   \author{M. E. Shultz
          \inst{1}
          \and
          I.\ Berry
          \inst{2}
          \and
          D.\ Bohlender
          \inst{3}
          \and
          G.\ Catanzaro
          \inst{4}
          \and
          M.\ Giarrusso
          \inst{4}
          \and
          R.\ Klement
          \inst{1}
          \and
          J.\ Labadie-Bartz
          \inst{5}
          \and
           F.\ Leone
           \inst{4,6}
          \and
           P.\ Leto
           \inst{4}
          \and
          C.\ Neiner
          \inst{7}
          \and
          S.\ P.\ Owocki
          \inst{8}
          \and
           Th.\ Rivinius
           \inst{1}
          \and
           A.\ ud-Doula
           \inst{9}
          \and
           G.\ A.\ Wade
           \inst{10}
          }

   \institute{
         ESO - European Organisation for Astronomical Research in the Southern Hemisphere, Casilla 19001, Santiago 19, Chile\\
                  \and
                  Institute for Astronomy, University of Hawai'i at Mānoa, Honolulu, HI 96822 USA\\
         \and
         National Research Council of Canada, Herzberg Astronomy and Astrophysics Research Centre, 5071 West Saanich Road, Victoria, BC V9E 2E7\\
         \and
         INAF - Osservatorio Astrofisico di Catania, Via S. Sofia 78, 95123 Catania, Italy\\
        \and
         Instituto de Astronomia, Geo\'isica e Ci\^encias Atmosf\'ericas, Universidade de S\=ao Paulo, Rua do Mat\=ao 1226, Cidade Universit\'aria,\\ B-05508-900 S\=ao Paulo, SP, Brazil\\
         \and
         Dipartimento di Fisica e Astronomia, Sezione Astrofisica, Universitá di Catania, Via S. Sofia 78, I-95123 Catania, Italy\\
         \and
         1 LESIA, Paris Observatory, PSL University, CNRS, Sorbonne Universit\'e, Universit\'e de Paris, 5 place Jules Janssen, 92195 Meudon, France\\
         \and
                      Department of Physics and Astronomy, University of Delaware, 217 Sharp Lab, Newark, Delaware, 19716, USA\\
         \and
         Department of Physics, Penn State Scranton, Dunmore, PA 18512, USA\\
         \and
         Department of Physics and Space Science, Royal Military College of Canada, Kingston, Ontario K7K 7B4, Canada\\
             }

   \date{Received September 15, 1996; accepted March 16, 1997}
 
  \abstract
   {The nearby B-type star $\rho$\,Oph\,A was recently identified as a rapidly rotating magnetic B-type star with variable radio and X-ray emission consistent with a magnetospheric origin.}
   {We aim to refine constraints on $\rho$\,Oph\,A's magnetic field using high-resolution spectropolarimetry, and to scrutinize its spectrum for indications of the expected H$\alpha$ emission.}
   {We present a high-resolution spectropolarimetric time series obtained with ESPaDOnS, which we use to perform a magnetic analysis using least-squares deconvolution (LSD). The star's rotation period is revisited using Kepler-2 space photometry.}
   {We find that $\rho$\,Oph\,A is a spectroscopic binary consisting of two B-type stars with masses of about 8 and 10 \msun~on a slightly eccentric 88-day orbit, with the magnetic field being associated with the smaller Ab component. This leads to $\rho$\,Oph\,Ab's 4~kG surface magnetic dipole being approximately twice as strong as the previously reported 2~kG. The oblique rotator model derived from the longitudinal magnetic field curve agrees well with the LSD Stokes $V$ profiles, indicating that the magnetic field is likely to be very nearly dipolar. The orbital and rotational axes appear to be aligned. We report for the first time $\rho$\,Oph\,Ab's magnetospheric H$\alpha$ emission, which is consistent with an origin in a centrifugal magnetosphere. We also demonstrate that $\rho$\,Oph\,Ab's light curve can be recovered from the previously reported {\em Kepler-2} light curve of $\rho$\,Oph\,C, and that it demonstrates prominent magnetospheric eclipses similar to those of $\sigma$\,Ori\,E, and which can be reproduced using a dipolar Rigidly Rotating Magnetosphere model. Both H$\alpha$ and the light curve are indicative of a strongly asymmetric magnetosphere.}
   {All indications are that $\rho$\,Oph\,Ab's magnetic field is essentially dipolar, meaning that contributions from higher-order multipoles probably cannot explain the strong asymmetry in the magnetosphere. Only two other stars show comparable degrees of asymmetry, both of which are also close binaries, suggesting that binarity can affect the magnetospheric plasma distributions.}


   \maketitle

\section{Introduction}

Due to their unique magnetic properties and the rich variety of observational phenomena associated with their magnetospheres, magnetic early B-type stars have emerged over the past decade as powerful laboratories for magnetospheric physics. Despite the rarity of magnetic fields in the overall hot star population \citep[an incidence rate of about 10\%;][]{2017MNRAS.465.2432G,2019MNRAS.483.2300S}, magnetic hot stars are often easily identified due to their striking chemical peculiarities. Being much more luminous objects than cool stars, observationally expensive spectropolarimetric datasets can be, and have been, collected for hundreds of objects \citep[e.g.][]{2018MNRAS.475.5144S,2019MNRAS.483.2300S,2019MNRAS.489.5669P,2021MNRAS.504.5813M,2021AstBu..76..163R}. 

Unlike cool star magnetic fields, which exhibit intrinsic variability due to their magnetic activity cycles \citep[e.g.][]{2016MNRAS.459.4325M}, the surface magnetic fields of hot stars are not supported by a contemporaneous dynamo, but are thought to be `fossils' imprinted in the radiative envelope at a previous evolutionary stage such as e.g.\ binary mergers \citep[][]{2004Natur.431..819B,2010ApJ...724L..34D,2019Natur.574..211S,2021MNRAS.504.2474K,2023Sci...381..761S,2024Sci...384..214F}. These fossil magnetic fields are absolutely stable over human timescales, exhibiting no variability over at least thousands of rotational cycles \citep{2018MNRAS.475.5144S}, although they exhibit a slow decline over evolutionary timescales \citep{land2007,land2008,2016A&A...592A..84F,2019MNRAS.490..274S,2019MNRAS.483.3127S}. Finally, and again in contrast to cool stars whose magnetic field exhibit significant structure at both small and large scales \citep[e.g.][]{2020A&A...635A.142K}, hot star fossil fields are generally geometrically simple, with most of the population being well approximated by dipoles tilted with respect to the rotation axis \citep[e.g.][]{2018MNRAS.475.5144S,2019A&A...621A..47K}. 

The stability and simplicity of hot star magnetic fields provides an easily modeled, time-invariant background against which to evaluate the behaviour of their magnetospheres, which are fed by their radiatively driven winds \citep[][]{1984NASCP2349..487B,brown1987,1988ESASP.281a.339S,shorebrown1990,1990ApJ...365..665S}. Magnetic confinement of the wind plasma leads to wind flows being channeled towards the magnetic equator, where colliding flows from opposite co-latitudes stall and settle into a cool, dense plasma within the equatorial plane \citep[e.g.][]{ud2002}. Corotation of the plasma induced by the Lorentz force leads to rotational modulation of the associated photometric eclipses produced by plasma occluding the star, a phenomenon which enabled discovery of the first stellar magnetosphere around $\sigma$\,Ori\,E by \cite{lb1978}. If the star's surface rotational velocity is negligible relative to the ability of the magnetic field to confine the wind, the plasma is pulled back to the star by gravity, leading to the formation of a `dynamical magnetosphere' \citep[DM;][]{petit2013}. If the star is rapidly rotating, the extreme centrifugal stress on the confined plasma introduced by corotation prevents gravitational infall, leading to the formation of a dense `centrifugal magnetosphere' region \citep[CM;][]{petit2013}. Since the magnetic field itself is stable over rotational timescales, DMs and CMs can be respectively modeled in a relatively straightforward fashion using the Analytic Dynamical Magnetosphere \citep[ADM;][]{2016MNRAS.462.3830O} and Rigidly Rotating Magnetosphere \citep[RRM;][]{town2005c} models. Photometric ADM modeling has been performed for magnetic O-type stars \citep{2020MNRAS.492.1199M}, and photometric RRM models \citep{town2008,2022MNRAS.511.4815B} have been applied to several magnetic B-type stars, e.g.\ $\sigma$\,Ori\,E and HD\,37776 \citep{town2005b,town2013,2015MNRAS.451.2015O,2022A&A...659A..37K,2023MNRAS.523.6371B}. 

In addition to photometric eclipsing, magnetospheres are often detectable via a variety of diagnostics: they are commonly seen in X-ray \citep[e.g.][]{2014ApJS..215...10N}, H$\alpha$ \citep[e.g.][]{petit2013,2019MNRAS.490..274S}, radio gyrosynchrotron \citep[e.g.][]{1993A&A...268..667L,2021MNRAS.507.1979L,2022MNRAS.513.1429S}, and beamed electron-cyclotron maser emission \citep[ECME, e.g.][]{2000A&A...362..281T,2022ApJ...925..125D,2022MNRAS.517.5756D}. Correlations between H$\alpha$ and incoherent radio emission strengths \citep{2022MNRAS.513.1429S}, and of both forms of emission with rotation and magnetic field strength \citep{2021MNRAS.507.1979L,2022MNRAS.513.1429S}, link them to the same Centrifugal Breakout (CBO) mechanism , in which the wind feeds the magnetosphere up to the capacity of the magnetic field lines to hold the magnetically confined plasma, at which point the plasma is ejected away from the star in explosive magnetic reconnection events \citep{1984AA...138..421H,ud2008,2020MNRAS.499.5366O,2022MNRAS.513.1449O}. While scaling laws based on CBO apparently explain both H$\alpha$ and gyrosynchrotron emission, the X-ray overluminosity of magnetic hot stars is instead believed to originate in magnetically confined wind shocks \citep{bm1997,ud2014}. A scaling law based on magnetically confined wind shocks explains most of the variance in the X-ray luminosity of the magnetic hot star population \citep{2014ApJS..215...10N}. However, \cite{2017MNRAS.467.2820L,2018MNRAS.476..562L} reported tentative detections of non-thermal X-rays, which they postulated may be a result of X-ray aurorae arising due to bremsstrahlung emission from collision with the stellar atmosphere of the same non-thermal electrons responsible for gyrosynchrotron and ECME.

The possibility of X-ray aurorae was given further support by the detection of what appears to be a high-amplitude rotational modulation of the X-ray light curve of the magnetic B-type star $\rho$\,Oph\,A (HD\,147933, B2/3\,V). This modulation was first reported by \cite{2017A&A...602A..92P}, with the first magnetic detection (using low-resolution FORS2 spectropolarimetry) presented by \cite{2018A&A...610L...3P}. \cite{2020MNRAS.493.4657L} presented follow-up FORS2 spectropolarimetry of the object, together with a comprehensive radio-frequency time series that revealed both rotationally modulated gyrosynchrotron radiation and ECME. Despite being an extremely rapid rotator (with a rotation period $P_{\rm rot} \sim 0.74~{\rm d}$) with a moderately strong magnetic field (a surface magnetic dipole strength $B_{\rm d} \sim$ 2.7 kG), i.e.\ exactly the properties that would normally imply H$\alpha$ emission from its CM, no H$\alpha$ emission from this object has ever been reported. \cite{2020MNRAS.493.4657L} also reported an apparent conflict between the cluster age and the spindown age inferred from the magnetospheric braking timescale \citep[e.g.][]{ud2009,petit2013}, with a maximum spindown age of $\rho$\,Oph\,A of between 10 and 40 Myr, as compared to the age of the $\rho$ Ophiuchus group of 5-10 Myr found by \cite{2016A&A...592A..88P}. 

In this paper we present the first high-resolution spectropolarimetric time series for $\rho$\,Oph\,A, with the intent of establishing tight constraints on the magnetic field strength and geometry of this object, and probing its H$\alpha$ magnetosphere. The observations are discussed in Sect.\ \ref{sect:obs}. In Sect.\ \ref{sect:mult} we demonstrate that $\rho$\,Oph\,A is a previously unidentified spectroscopic binary, and determine the orbital period and elements. Atmospheric and fundamental parameters for the two stars are derived in Sect.\ \ref{sect:stellarpars}. The rotational period is re-examined in Sect.\ \ref{sect:rot}, where we find that the {\em K2} light curve of $\rho$\,Oph\,C also contains a contribution from the A component, which shows clear magnetospheric eclipses. In Sect.\ \ref{sect:magfld} we evaluate the magnetic field. The magnetosphere is probed using both H$\alpha$ and photometric modeling in Sect.\ \ref{sect:magnetospheres}, where we present for the first time $\rho$\,Oph\,A's H$\alpha$ emission properties, and demonstrate that its light curve is consistent with RRM predictions. The implications of our results are discussed in Sect.\ \ref{sect:discuss}. The main results and conclusions of the paper are summarized in Sect.\ \ref{sect:summary}. 

\section{Observations}\label{sect:obs}

\subsection{Spectropolarimetry}\label{subsect:esp}

As its name implies, the Echelle SpectroPolarimetric Device for the Observation of Stars (ESPaDOnS) is a high-resolution ($R \sim 65,000$ at 500~nm) \'echelle spectropolarimeter, covering the spectral region between about 370 and 1050 nm across 40 overlapping orders. It has been mounted at the Cassegrain focus of the 3.6~m Canada-France-Hawaii Telescope (CFHT) since 2004, and has been the primary work-horse for stellar magnetometry for almost two decades. It obtains spectropolarimetric sequences consisting of 4 differently polarized sub-exposures, yielding 4 unpolarized (Stokes $I$) spectra, and one polarized spectrum in either circular (Stokes $V$) or linear (Stokes $Q$ and $U$) polarization. By combining the sub-exposures in such a fashion that intrinsic source polarization cancels out, a diagnostic null $N$ spectrum is also obtained; this enables the noise level to be precisely quantified, while providing a check for normal instrument operation. A clone of ESPaDOnS, Narval, was mounted at the 2~m Bernard Lyot Telescope\footnote{Narval has recently been replaced by NeoNarval.}. The properties of the two instruments, their data reduction using the Libre-ESpRIT package \citep{d1997} and the CFHT Upena pipeline, and the excellent agreement in their results, were described in detail by \cite{2016MNRAS.456....2W} in the context of the MiMeS Large Programs. 

Between 18/05 and 28/05/2021 we obtained 12 ESPaDOnS Stokes $V$ sequences of $\rho$\,Oph\,A. Each sequence used a 225 sec sub-exposure time, for a total exposure time of 900 sec. These observations were spaced so as to evenly sample the 0.74~d rotation period determined by \cite{2020MNRAS.493.4657L}. One sequence had an unacceptably low maximum signal-to-noise $S/N$ per 1.8 \kms~pixel in Stokes $I$ of 621, and this rotation phase was re-observed; discarding this observation, the mean peak $S/N$ of the dataset is 1490, with a range from 1370 to 1637. The high target $S/N$, and the exposure times required to reach it, were guided by the rotational broadening and surface magnetic field strength reported by \cite{2020MNRAS.493.4657L}.

$\rho$\,Oph\,A was observed 8 times with Narval on 10/07/2014, using a sub-exposure time of 100 sec. The observations were collected over a 1.2-hour time-span, and are therefore at essentially the same rotational phase (spanning about 0.06 of a rotational cycle). They were therefore co-added, yielding a single spectrum with a maximum $S/N$ of 895. 

All ESPaDOnS and Narval observations were normalized using standard order-by-order polynomial normalization \citep[e.g.][]{2018MNRAS.475.5144S}. 

\subsection{Spectroscopy}\label{subsect:spec}

In addition to the spectropolarimetry, various spectra were obtained for $\rho$\,Oph\,A using a variety of instruments. The majority of these observations were reported by \cite{2020MNRAS.493.4657L}, and are included in the current work. These include 2 unpolarized ESPaDOnS spectra, 3 spectra obtained with the ESO UV-Echelle Visible Spectrograph \citep[UVES;][]{2000SPIE.4008..534D}, 20 spectra obtained with Catania Astrophysical Observatory Spectropolarimeter \citep[CAOS;][]{2016AJ....151..116L}, and 4 spectra obtained with the High Accuracy Radial velocity Planet Searcher for the Northern hemisphere \citep[HARPS-N;][]{2012SPIE.8446E..1VC}.

We have also obtained new spectra using the Dominion Astrophysical Observatory (DAO) 1.2m telescope and McKellar spectrograph operating robotically. The detector was the $2048 \times 4088$ pixel SITe-4 CCD with 15~$\mu$m square pixels. The spectrograph was configured with the IS32R imaging slicer, the 32 inch camera, and the 1200H grating. The central wavelength was set near 6600~\AA. This configuration produces a dispersion of 0.15 \AA~/ pixel, with a resolution $\lambda/\Delta\lambda \sim 17000$. Processing was carried out with a standard pipeline for single slit spectra. The reduction  sequence included bias subtraction, flat-fielding, optimal extraction of the spectrum, wavelength calibration, and a correction to heliocentric velocities.  

\subsection{Time Domain Photometry}

While no modern time-series photometry exists for $\rho$\,Oph\,A specifically, \cite{2018AJ....155..196R} determined the rotational period of the nearby $\rho$\,Oph\,C using the {\em K2} light curve, using an extraction mask that included light from $\rho$\,Oph\,A. The stars are separated by only $2.5^{\prime}$, meaning that stray light from $\rho$\,Oph\,A is included in the extraction mask used to obtain $\rho$\,Oph\,C's light curve (as $\rho$\,Oph\,C is a bright star, the central pixels were saturated, necessitating a relatively large mask).

The {\em K2} mission was an extension of the NASA {\em Kepler} mission \citep{2014PASP..126..398H}. The {\em Kepler} space telescope was a $\mu$mag precision photometer launched in 2009 with the goal of obtaining high-cadence, long-duration time series photometry in order to detect exoplanets via the transit method. When two of the reaction wheels failed, the {\em K2} mission extension was developed. {\em K2} used solar wind pressure to stabilize the satellite on a given field of view along the ecliptic for periods of about 3 months. $\rho$\,Oph\,C was observed during campaign 2, between 24/08 and 10/11/2014. We obtained these data from the Mikulski Archive for Space Telescopes (MAST). 

\section{Multiplicity of $\rho$ Oph A}\label{sect:mult}

   \begin{figure}
   \centering
   \includegraphics[width=0.45\textwidth]{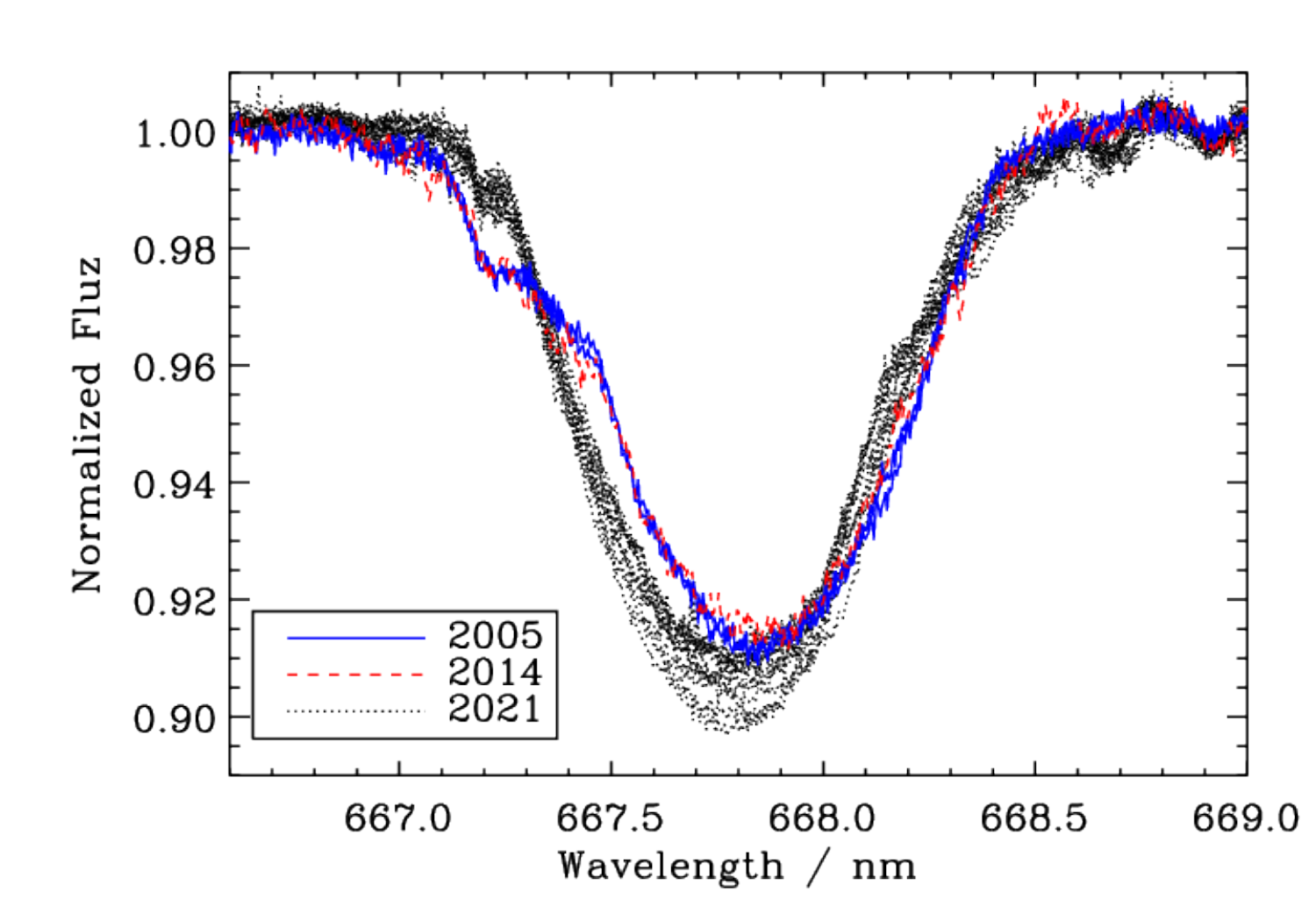} 
      \caption[]{He~{\sc i}~667.8~nm line of $\rho$ Oph A as measured in 2005, 2014, and 2021 from ESPaDOnS and Narval spectra.}
         \label{rhoOphA_HeI6678}
   \end{figure}

   \begin{figure}
   \centering
   \includegraphics[width=0.45\textwidth]{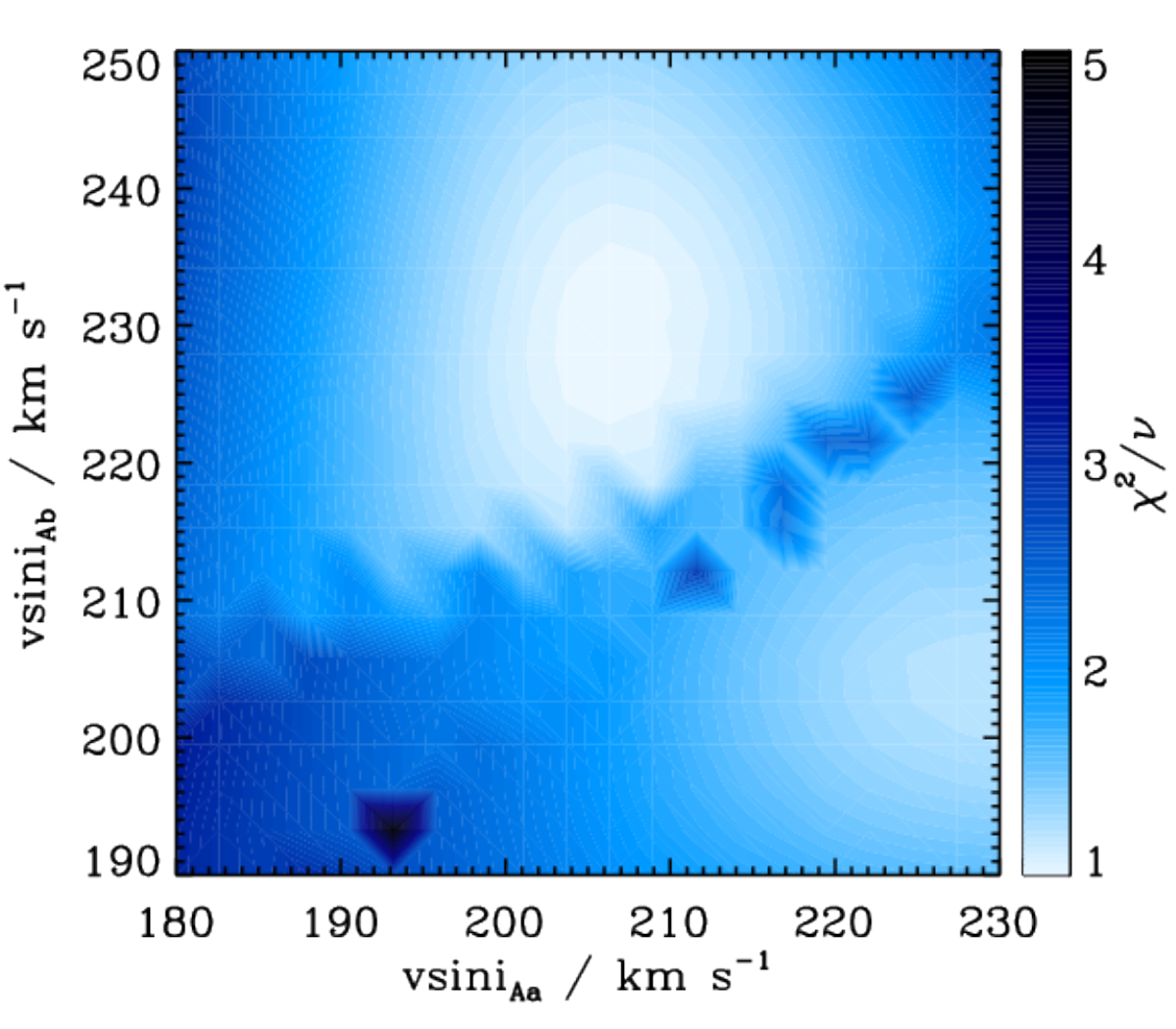} 
      \caption[]{Reduced $\chi^2$ landscape for the \vsini~determination of the components of $\rho$ Oph A.}
         \label{vsini_gof}
   \end{figure}

   \begin{figure}
   \centering
   \includegraphics[width=0.4\textwidth,trim=50 50 100 0]{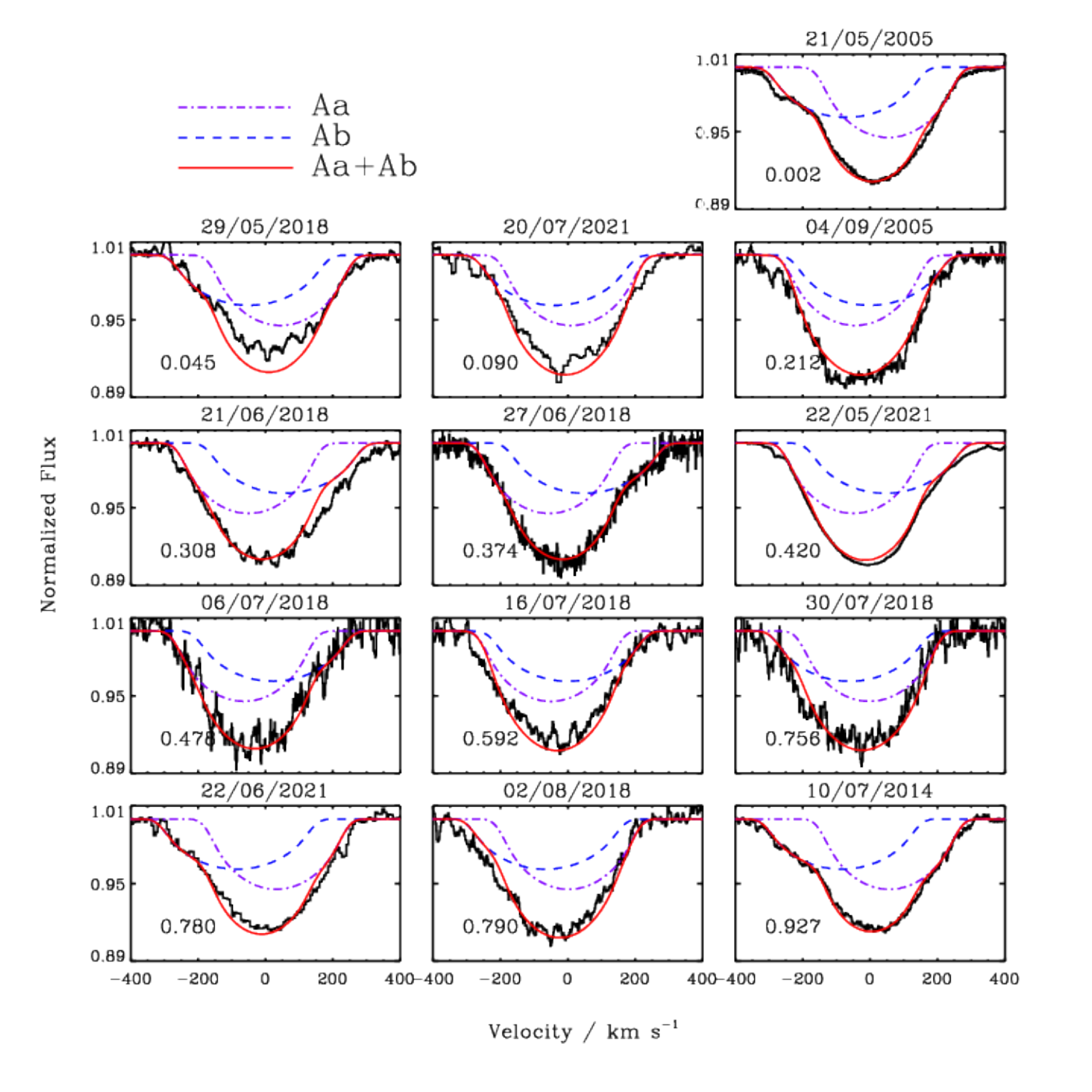} 
      \caption[]{Binary fits to the He~{\sc i}~667.8~nm line profiles of $\rho$ Oph A, arranged in order of increasing orbital phase from top to bottom right (see labels in bottom left of each panel).}
         \label{rhoOphA_HeI6678_binary_fit}
   \end{figure}

Comparison of the line profiles from previous Narval and ESPaDOnS observations to those obtained in the 2021 run reveals systematic differences, as can be seen in Fig.\ \ref{rhoOphA_HeI6678} for the He~{\sc i}~667.8~nm line, chosen for being a strong and isolated feature. The 2021 observations cover the entire 0.74~d rotational period, therefore the discrepancies cannot be ascribed to rotation. Indeed, there is clear variability in the core of the line, which is almost certainly rotational in origin, but these variations at no point reconcile the new dataset with previous observations.

The most straightforward explanation for the long-term variations is binarity. $\rho$ Oph AB is a well-known visual binary, for which \cite{2007BaltA..16..435N} found an orbital period of $2398 \pm 326$~yr. The components are separated by $4.25 \pm 0.8 \arcsec$, which is larger than the ESPaDOnS pinhole, therefore $\rho$ Oph B should not have affected the measurements. The long orbital period of $\rho$\,Oph\,AB further rules out stray light from $\rho$ Oph B as the origin of the line profile variations, given the much shorter time-frame over which radial velocity variations are apparent. We therefore infer that $\rho$\,Oph\,A is a previously unidentified two-lined spectroscopic binary (SB2). 

\subsection{Radial velocities}\label{subsect:rv}

Measuring radial velocities (RVs) from the $\rho$ Oph A components requires an unusual degree of care. The typical procedure is to fit a two-component parameterized model to the line profile \citep[e.g.][]{2017MNRAS.465.2432G} in which \vsini, RVs, and the equivalent width (EW) ratio are free parameters. However, this requires that the line profiles of the two stars are clearly distinguishable, i.e.\ that the reflex motion is comparable to or larger than the velocity broadening. As can be seen in Fig.\ \ref{rhoOphA_HeI6678} this is far from the case: both components are apparently very broad, with \vsini~at least 200 \kms, and a RV amplitude small enough that the components are always strongly blended. The usual approach therefore cannot be relied upon.

Instead, RVs, \vsini, and EW ratios were determined by applying a grid of model line profiles, generated via standard disk integration, to the entire dataset. The disk integration model is based on the one described by \cite{petit2012a}, originally intended for modeling Least-Squares Deconvolution mean line profiles, with local line profiles treated as Gaussians and then combined using a given \vsini~and limb darkening coefficient. Since the model does not include radiative transfer or make any assumptions about surface abundances, the equivalent width of the resulting line profiles is an arbitrary scaling parameter. Since there is no evidence for large RV variations on timescales of less than 10 days (the duration of the 2021 ESPaDOnS observing run), mean spectra created with 10-day bins were used. 

At each point on the grid, the \vsini~and EW ratio of the two components were kept fixed. Based on analysis of the 2005 ESPaDOnS spectrum -- in which the two components are relatively well separated -- the \vsini~of the stronger component (which we denote Aa) was varied between 180 and 230 \kms, while the \vsini~of the second, weaker component (Ab) was varied between 190 and 250 \kms. The EW ratio (i.e. the ratio of the EW of the stronger component to the total EW) was varied between 0.51 and 0.9. For each combination of \vsini~and EW ratio, the RVs of each component were allowed to vary between $\pm$150 \kms, until a global $\chi^2$ minimum was found for the full dataset. As an initial guess, each component was assigned the RV determined from the centre-of-gravity of the line. 

Fig.\ \ref{vsini_gof} shows the $\chi^2$ landscape, with the $\chi^2$ normalized to unity at the global minimum by subtracting the minimum value. The best-fit solution yields \vsini$_{\rm Aa} = 206 \pm 5$~\kms, \vsini$_{\rm Ab} = 227 \pm 5$~\kms, and an EW ratio of 0.55. As can be seen in Fig.\ \ref{rhoOphA_HeI6678_binary_fit}, the resulting composite profiles provide a good fit to all observations. The RVs measured in this process are provided in Table \ref{table:rv}. Uncertainties were estimated to be around 10 \kms, given the low amplitude of the RVs relative to \vsini. 

\subsection{Orbit}\label{subsect:orbit}

\begin{table}
\centering
\caption[]{Orbital parameters for the $\rho$ Oph A system.}
\label{table:orbit}
\begin{tabular}{l c}
\hline\hline
Parameter & Value \\
\hline
$P_{\rm orb} / {\rm d}$ & 88.00(2) \\
$T_0 / {\rm HJD}$ & 2453511.8(7) \\
$e$ & $0.13 \pm 0.05$ \\
$\gamma / {\rm km~s^{-1}}$ & $-16 \pm 2$ \\
$K_{\rm Aa} / {\rm km~s^{-1}}$ & $57 \pm 5$ \\
$K_{\rm Ab} / {\rm km~s^{-1}}$ & $67 \pm 5$ \\
$\omega / ^\circ$ & $32 \pm 4$ \\
$q$ & $1.2 \pm 0.2$ \\
$M_{\rm Aa} \sin^3{i} / {\rm M_\odot}$ & $9.1 \pm 2.3$ \\
$M_{\rm Ab} \sin^3{i} / {\rm M_\odot}$ & $7.8 \pm 2.1$ \\
$M_{\rm (Aa + Ab)} \sin^3{i} / {\rm M_\odot}$ & $17 \pm 6$ \\
$a\sin{i} / {\rm AU}$ & $0.99 \pm 0.07$ \\
$i_{\rm orb} / ^\circ$ & $69 \pm 11$ \\
$a / {\rm AU}$ & $1.1 \pm 0.2$ \\
\hline
\hline\hline
\end{tabular}
\end{table}

   \begin{figure}
   \centering
   \includegraphics[width=0.45\textwidth]{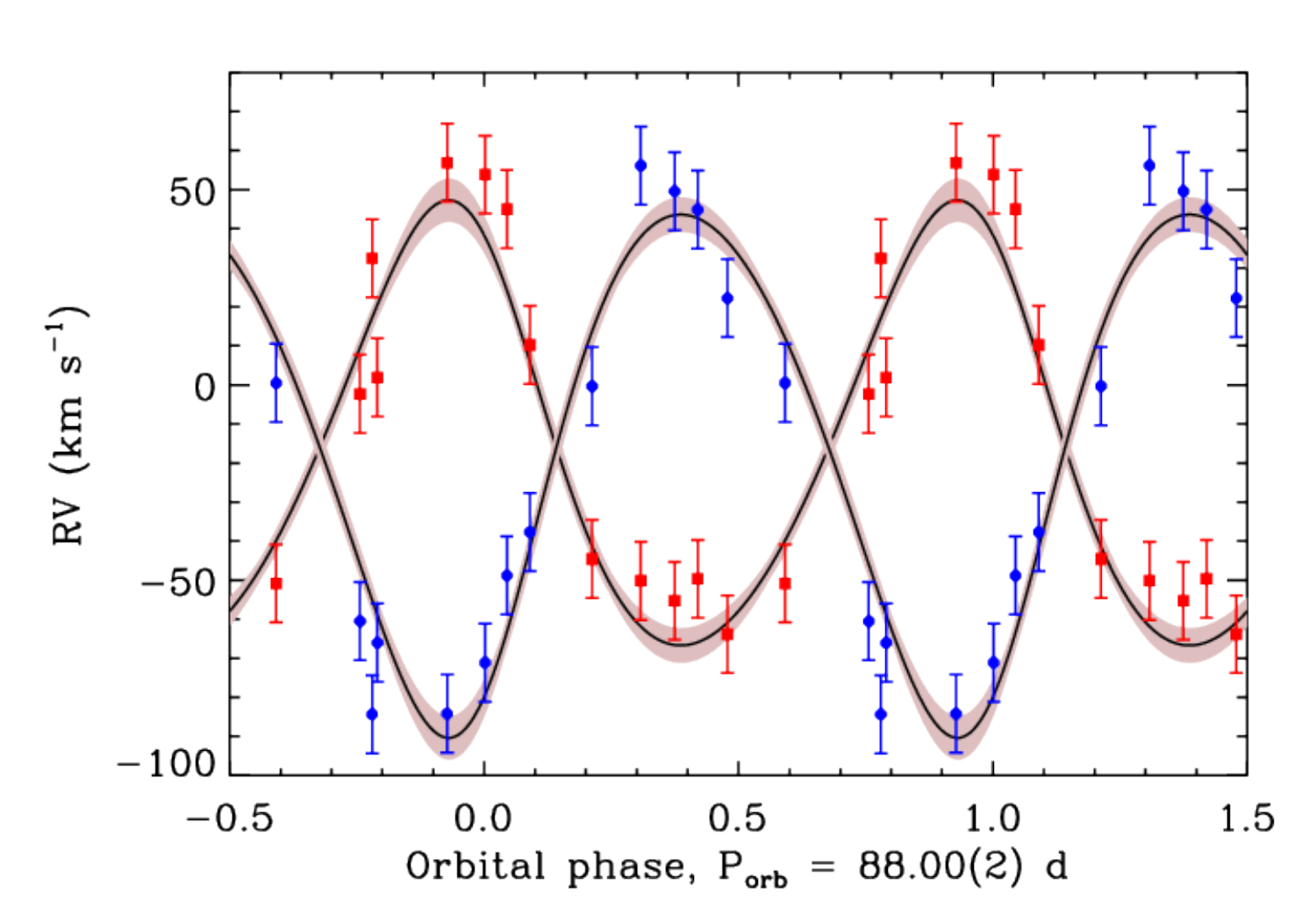} 
      \caption[]{Radial velocity measurements of $\rho$ Oph Aa (red squares) and Ab (blue circles), folded with the orbital period. Curves and shaded regions represent the orbital model and uncertainties.}
         \label{rhoOphA_orb}
   \end{figure}

In order to determine the orbital period, the RVs of the Aa and Ab components were separately analyzed using the Lomb-Scargle {\sc idl} program {\sc periodogram.pro}\footnote{\url{https://hesperia.gsfc.nasa.gov/ssw/gen/idl/util/periodogram.pro}} \citep{1986ApJ...302..757H}, as well as the standard Fourier analysis package {\sc period04}\footnote{\url{https://www.univie.ac.at/tops/Period04/}} \citep{2005CoAst.146...53L}. This yielded maximum power at 88.1(1)~d for the Aa RVs, and 87.9(2)~d for the Ab RVs, where the numbers in parentheses here and elsewhere in the text indicate analytic 1$\sigma$ uncertainties in the least significant digit as inferred by {\sc period04} using the formula given by \cite{1976fats.book.....B}, which accounts for the number and time-span of observations, the amplitude of the signal, and the measurement error. We adopt 88.0(2)~d as the most likely orbital period. The RVs are shown folded with this period in Fig.\ \ref{rhoOphA_orb}.

Orbital parameters were determined using the Markov Chain Monte Carlo (MCMC) algorithm described by \cite{2018MNRAS.475..839S,2019MNRAS.482.3950S,2021MNRAS.504.3203S}. The resulting orbital model is shown in Fig.\ \ref{rhoOphA_orb}, and the orbital elements are given in Table \ref{table:orbit}. The orbit is mildly eccentric ($e = 0.13 \pm 0.05$), with a mass ratio $q = M_{\rm Aa}/M_{\rm Ab} = 1.2 \pm 0.2$ consistent with the similar strengths of the contributions of the two stars to the He~{\sc i}~667.8~nm line. The projected masses of the two components, about 8 and 9 \msun, are furthermore consistent with the SIMBAD B2/3\,V spectral type assigned under the assumption that $\rho$ Oph A is a single star, and with the \teff~determined by \cite{2020MNRAS.493.4657L} under the same assumption. Since the masses inferred from the mass function increase with decreasing orbital inclination, this indicates that the orbital plane is likely at high inclination from the line-of-sight, and that the true semi-major axis is similar to the projected value of about 1 AU. 

\section{Stellar parameters}\label{sect:stellarpars}

\begin{table}
\centering
\caption[]{Stellar parameters. Reference key: $^a$\protect\cite{1994ApJ...424..772F}; $^b$\protect\cite{2002A&A...384..180F}}
\label{table:stellar_pars}
\begin{tabular}{l c c}
\hline\hline
Parameter & Aa & Ab \\
\hline
\\
$\pi / {\rm mas}$ & \multicolumn{2}{c}{$7.19 \pm 0.17$} \\
$d / {\rm pc}$ & \multicolumn{2}{c}{$139 \pm 3$} \\
$E(B-V) / {\rm mag}$ & \multicolumn{2}{c}{$0.59 \pm 0.02^a$} \\
$V / {\rm mag}$ & \multicolumn{2}{c}{$5.05^b$} \\
$BC / {\rm mag}$ & $-2.24 \pm 0.08$ & $-1.85 \pm 0.10$ \\
$M_{\rm bol} / {\rm mag}$ & $-4.36 \pm 0.22$ & $-2.96 \pm 0.32$ \\
$\log{L_{\rm bol} / {\rm L_\odot}}$ & $3.63 \pm 0.09$ & $3.02 \pm 0.14$ \\
$T_{\rm eff} / {\rm kK}$ & $23 \pm 1$ & $19 \pm 1$ \\
$\log{g / {\rm cm~s^{-2}}}$ & $4.25 \pm 0.25$ & $4.25 \pm 0.25$ \\
\vsini~$ / {\rm km~s^{-1}}$ & $206 \pm 5$ & $227 \pm 5$ \\
$R_* / {\rm R_\odot}$ & $4.2 \pm 0.3$ & $3.1 \pm 0.2$ \\
$M_* / {\rm M_\odot}$ & $8.4 \pm 0.5$ & $6.0 \pm 0.5$ \\
${\rm Age / Myr}$ & \multicolumn{2}{c}{$15^{+6}_{-6}$} \\
\\
\hline\hline
\end{tabular}
\end{table}

The atmospheric and fundamental parameters of $\rho$ Oph A were previously determined by \cite{2020MNRAS.493.4657L} under the assumption of a single star, and therefore must be revisited with the discovery that the star is an SB2. The stellar parameters determined here are summarized in Table \ref{table:stellar_pars}. 

\subsection{Atmospheric parameters}\label{subsec:atmos}

   \begin{figure*}[hbt!]
   \centering
   \includegraphics[width=0.95\textwidth]{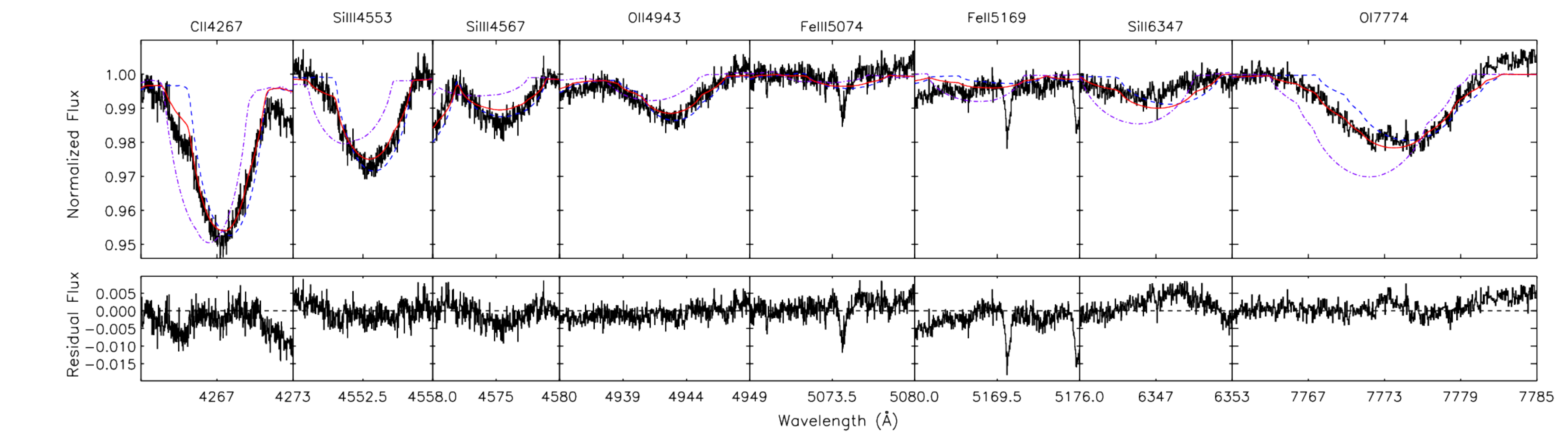} 
   \includegraphics[width=0.95\textwidth]{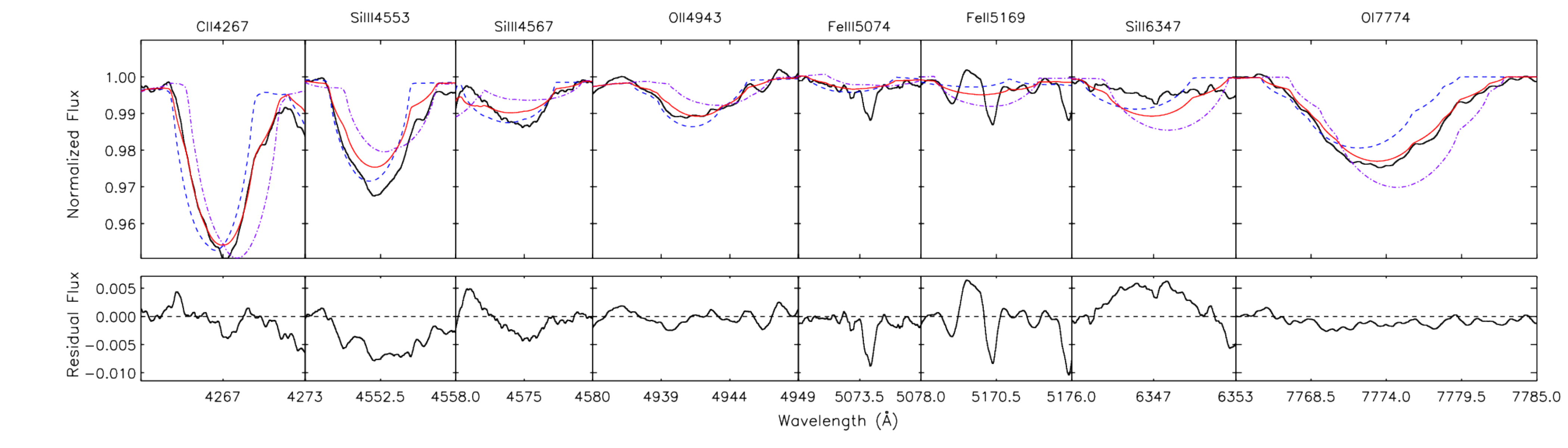}
   \includegraphics[width=0.95\textwidth]{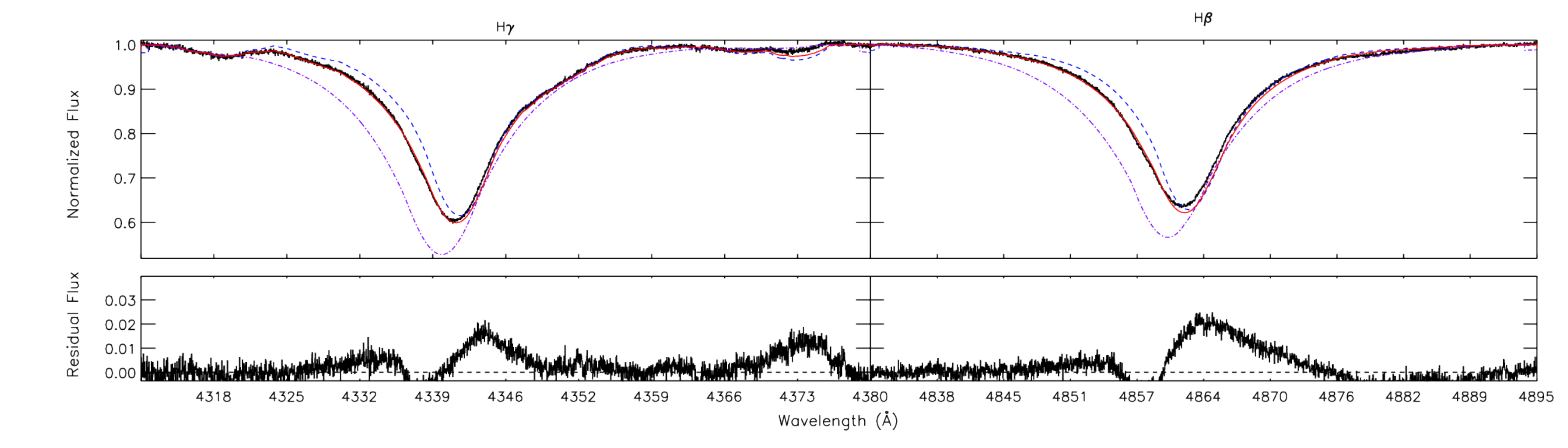}
      \caption[]{Best-fit synthetic {\sc tlusty} spectrum for $\rho$ Oph A (Aa: dashed blue; Ab: dot-dashed purple; combined: red) to the mean 2005 ESPaDOnS spectrum (top), the 2021 mean ESPaDOnS spectrum (middle), and in the vicinity of H$\beta$ and H$\gamma$ in 2015 (bottom). Note that the synthetic profiles of the individual components are not weighted by their relative luminosities.}
         \label{rhoOphA_met_line_fit}
   \end{figure*}

To determine the effective temperatures \teff~and surface gravities $\log{g}$ of $\rho$\,Oph\,Aab, we used the standard BSTAR2006 library of solar metallicity {\sc tlusty} non-local thermodynamic equilibrium synthetic spectra \citep{lanzhubeny2007}, adopting the grid-based goodness-of-fit approach used by \cite{2021MNRAS.504.3203S,2021MNRAS.504.4850S}. To determine \teff, a selection of metallic lines spanning a range of ionization levels (C~{\sc ii}, Si~{\sc ii}, Si~{\sc iii}, O~{\sc i}, O~{\sc iii}, Fe~{\sc ii}, and Fe~{\sc iii}) were used, while $\log{g}$ was determined by fitting the wings of the H$\beta$ and H$\gamma$ profiles (H$\alpha$ was not used due to clear magnetospheric emission, analyzed in greater detail in Sect.~\ref{sect:magnetospheres}.) The \vsini~and RVs of the two components were held fixed to the values determined in the analysis of the He~{\sc i}~667.8~nm line (see Sect.~\ref{subsect:rv}, Figs.\ \ref{vsini_gof} and \ref{rhoOphA_HeI6678_binary_fit}), while the radius ratio $R_{\rm Aa}/R_{\rm Ab}$ used to combine the spectra was varied as a free parameter between 0.1 and 10 at each point on the grid (the TLUSTY spectra already include continuum intensity, therefore $T_{\rm eff}$ is already accounted for). 

The top panels of Fig.\ \ref{rhoOphA_met_line_fit} show the best-fit model to the metallic lines in the 2005 ESPaDOnS spectrum, chosen as the RV separation is largest in this observation, while the middle panels show the best-fit model to the same lines in the 2021 ESPaDOnS spectrum, which has a smaller RV separation but a higher signal-to-noise ratio. There are unidentified sharp-lined features in Fe~{\sc ii}~5169, Fe~{\sc iii}~5074, and Si~{\sc ii}~6347; these are most likely Discrete Interstellar Bands (DIBs), and did not affect the fit. Both spectra were analyzed independently, and yielded the same results: $\rho$\,Oph\,Aa has \teff~$= 23 \pm 1$~kK, while $\rho$\,Oph\,Ab has \teff~$= 19 \pm 1$~kK, with $R_{\rm Aa}/R_{\rm Ab} = 1.2$ (slightly smaller than the value inferred below from the Kiel diagram; see Sect. \ref{subsubsect:rhoopha_fundpars}). These effective temperatures bracket the value determined by \cite{2020MNRAS.493.4657L}, as might be expected if the spectrum is analysed under the assumption that it originates in a single photosphere. The radius ratio is consistent with the EW ratio from the analysis of the He~{\sc i}~667.8~nm line, and with the similar masses inferred from orbital modelling (see Sect.~\ref{subsect:orbit} and Table \ref{table:orbit}). 

The bottom panels of Fig.\ \ref{rhoOphA_met_line_fit} show the fit to H$\beta$ and H$\gamma$ in 2015. The preferred solution for $\log{g}$ is 4.25 for Aa and 3.5 for Ab. Such a large discrepancy is unexpected in a close SB, as close binaries are expected to be coeval \citep{1994MNRAS.271..999B}\footnote{A merger origin could disrupt this assumption; this scenario is examined in greater detail in the Discussion.}. Since the less massive Ab component cannot be more evolved than Aa, the only other possibility is that while Aa is close to the zero-age main sequence (ZAMS), Ab is just beginning its contraction towards the ZAMS and is still very close to the birthline. This would imply that the system is at most a few hundred thousand years old, in sharp contrast to the cluster age determined by \cite{2016A&A...592A..88P} of 5--10 Myr, a measurement based upon both isochrone fitting to the full cluster population, and the fact that none of the nearby low-mass stars retain their protoplanetary disks. If true. this would be a remarkable result. However, the $\chi^2$ minimum for the surface gravity of Ab is quite broad, in contrast to the sharper minimum for Aa, and the fit with $\log{g} = 4.25$ for both stars is nearly indistinguishable in quality from the nominal best fit; it is this solution which is shown in the bottom panels of Fig.\ \ref{rhoOphA_met_line_fit}. Notably, the residuals are larger in H$\beta$ than in H$\gamma$, which may indicate that these lines are affected by magnetospheric emission, possibly explaining the anomalously low surface gravity preferred for $\rho$\,Oph\,Ab. In the end, we adopt $\log{g} = 4.25 \pm 0.25$ for both stars, as this provides an equivalently good fit to the data, and does not imply striking -- and implausible -- age differences between two presumably coeval stars.

\subsection{Distance to the $\rho$ Ophichus group}\label{subsubsect:rhooph_dist}

   \begin{figure}
   \centering
   \includegraphics[width=0.45\textwidth]{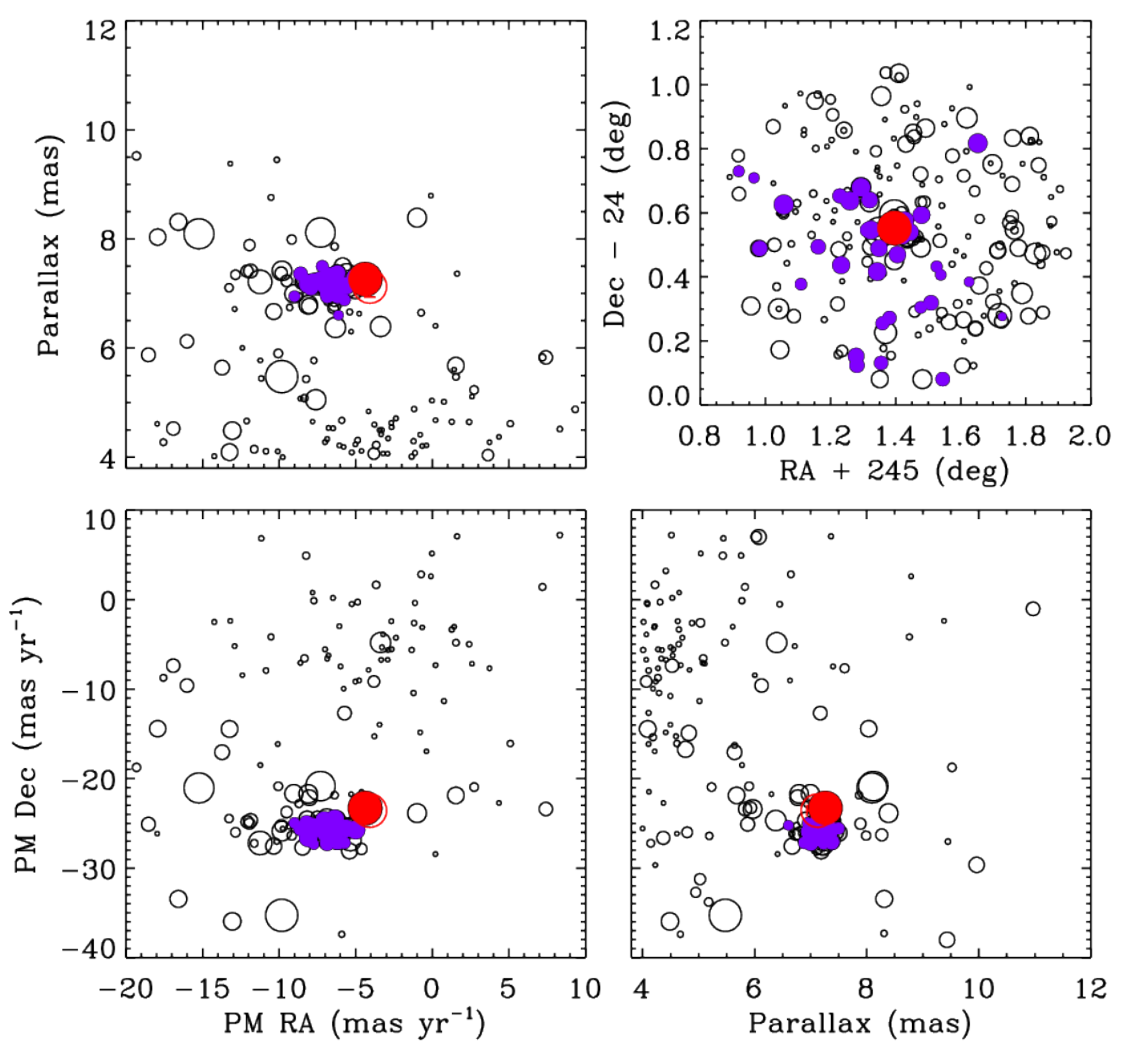}
      \caption[]{{\em Gaia} DR3 data proper motions and parallaxes for stars in the vicinity of $\rho$\,Oph\,A. Purple circles indicate high-probability members of the $\rho$\,Oph cluster. Red circles indicate $\rho$\,Oph\,A, with open circles indicating DR2 results for the same stars. The top right panel shows the distribution on the sky. Symbol size is inversely proportional to $G$ magnitude.}
         \label{rhooph_gaia}
   \end{figure}

As a first step in determining the fundamental parameters of $\rho$\,Oph\,Aa and Ab it is necessary to know their bolometric luminosities $L_{\rm bol}$. While {\em Gaia} Data Release 2 \citep[DR2;][]{2018A&A...616A...1G} and Data Release 3 \citep[DR3][]{2021A&A...649A...6G} parallaxes are available, these are not necessarily reliable. The renormalized unit weight error (RUWE) is 3.4, i.e.\ greater than unity; this is consistent with the star being a binary, and its parallax is therefore not reliable when treated with a single-star solution. 

Instead of using the parallaxes for the individual stars, the {\em Gaia} DR3 database was queried within 30$\arcmin$ of $\rho$\,Oph\,A's position, including all stars with a parallax between 4 and 10 mas. A previous clustering algorithm analysis of cluster membership based on DR2 data was performed by \cite{2019A&A...626A..80C}, who found a mean parallax of $7.2 \pm 0.2$~mas, and determined that $\rho$\,Oph\,A is a member. 

The Gaia astrometry is shown in Fig.\ \ref{rhooph_gaia}. Objects with RUWE greater than 1 were discarded. Cuts were then iteratively applied in proper motion and parallax, such that only those stars with values within 1$\sigma$ of the means were retained, with the means recalculated at each step in order to more precisely identify the true cluster value. $\rho$\,Oph\,A is indicated in Fig.\ \ref{rhooph_gaia}. It is unambiguously a member of the $\rho$\,Ophichus cluster. The mean parallax is $\pi = 7.19 \pm 0.17$~mas \citep[consistent with the value found by][]{2019A&A...626A..80C}, where the uncertainty reflects the standard deviation of the cluster members, yielding a distance $d = 139 \pm 3$~pc. 

\subsection{Reddening}

\cite{1994ApJ...424..772F} estimated $E(B-V) = 0.48$ in the region. \cite{2016A&A...592A..88P} used the DUST map, based on IRAS and COBE data, to derive $E(B-V) \sim 1$ in the vicinity of $\rho$ Ophiuchus, which gives an extinction $A_{\rm V} \sim 3$~mag assuming the usual reddening law $R_{\rm V} = 3.1$. However, the 3D STILISM tomographic dust map \citep[STructuring by Inversion the Local Interstellar Medium;][]{2014A&A...561A..91L,2017A&A...606A..65C} gives a much lower value of about $E(B-V) = 0.26$~mag at the distance to the cluster, with an asymptotic maximum of around 0.4 mag. 

While $\rho$\,Oph\,A does not have a full set of Johnson photometry, the nearby star $\rho$\,Oph\,C has measured UBVJHK magnitudes. Comparing the observed colours to the empirically determined intrinsic colours of young stars determined by \cite{2013ApJS..208....9P} for $\rho$\,Oph\,C's 16 kK \teff~yields $E(B-V) = 0.59 \pm 0.02$~mag, almost exactly in the middle of the values retrieved from the DUST and STILISM models. We therefore adopt this as the colour excess, yielding $A_{\rm V} = 1.83 \pm 0.06$~mag. 

\subsection{Fundamental parameters}\label{subsubsect:rhoopha_fundpars}

To determine the fundamental parameters of the $\rho$\,Oph\,A components, we used the binary Monte Carlo (MC) algorithm described by \cite{2019MNRAS.488...64P}. This takes the observed $V$ magnitude, extinction, distance, atmospheric parameters, mass ratio, and mass function of the system, and samples the Kiel diagram with points interpolated between evolutionary models, with probabilistic acceptance of a given solution if it accords with the priors within their uncertainties whilst maintaining coevality of the component stars. We used the rotating, solar metallicity Geneva evolutionary models calculated by \cite{ekstrom2012}. We used the \cite{nieva2013} bolometric correction $BC$, since Aa is not chemically peculiar and Ab does not show obvious signs of strong chemical peculiarity. 

The results are listed in Table \ref{table:stellar_pars}. Coevality is satisified if the stars are between 9 and 21 Myr in age, consistent within uncertainty with the upper bound of 10 Myr determined by \cite{2016A&A...592A..88P} for the $\rho$\,Ophiuchus association. The radii of the two components have a ratio $R_{\rm Aa}/R_{\rm Ab} = 1.36 \pm 0.13$, approximately consistent with the ratio determined from modelling the spectrum (Sect.\ \ref{subsec:atmos}), with bolometric luminosities $\log{L_{\rm bol}/{\rm L_\odot}} = 3.63 \pm 0.09$ and $3.02 \pm 0.14$ for Aa and Ab respectively. The masses of the two stars are about 2~\msun~lower than the nominal projected masses inferred from the orbital solution, although this is well within the uncertainties in the latter. 

\section{Rotational Period}\label{sect:rot}

   \begin{figure}
   \centering
   \includegraphics[width=0.45\textwidth]{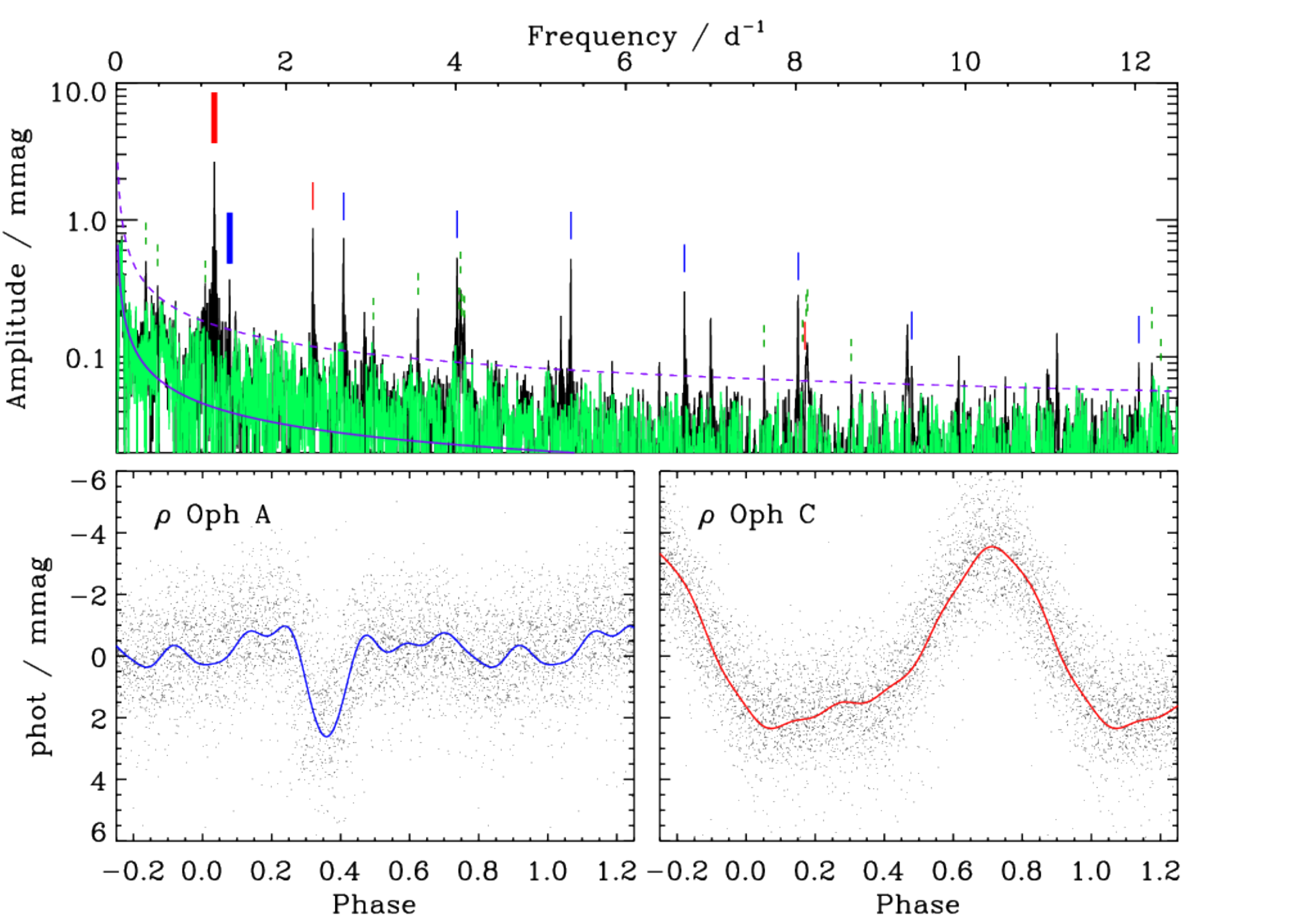}
      \caption[]{{\em Top}: Periodogram for the {\em K2} light curve of $\rho$\,Oph\,C. The rotational frequencies of $\rho$\,Oph\,Ab and C are indicated by long, thick blue and red dashes; harmonics of these frequencies are indicated by short dashes of the same colours. Significant frequencies that are not harmonics of either of the rotational frequencies are indicated by green dashes. The pre-whitened periodogram is shown in light green. The purple curve shows the noise model determined from the pre-whitened periodogram; the dashed curve indicates 4$\times$ the noise. {\em Bottom panels}: light curves of $\rho$\,Oph\,A ({\em left}) and C ({\em right}), folded with their rotational frequencies and pre-whitened with all frequencies unrelated to the star's rotation. Curves show harmonic models.}
         \label{rhooph_kepler}
   \end{figure}

\begin{table}
\centering
\caption[]{Rotational, magnetic, and magnetospheric parameters.}
\label{table:mag_pars}
\begin{tabular}{l c c}
\hline\hline
Parameter & Value & Uncertainty \\
\hline
$P_{\rm rot} / {\rm d}$ & 0.747326 & 0.000002 \\
$T_0 / {\rm HJD}$ & 2456849.71 & 0.06 \\[2.5 pt]
$i_{\rm rot} / ^\circ$ & 73 & $+4/-6$ \\[2.5 pt]
$v_{\rm eq} / {\rm km~s^{-1}}$ & 228 & $+13/-5$ \\
$R_{\rm pol}/R_{\rm eq}$ & 0.941 & $+0.002/-0.009$ \\
$W$ & 0.370 & $+0.031/-0.008$ \\
$R_{\rm K} / R_*$ & 1.93 & $+0.03/-0.08$ \\
\hline
$B0 / {\rm kG}$ & 0.20 & 0.04 \\
$B1 / {\rm kG}$ & 0.98 & 0.07 \\
$\Phi_1 / {\rm rad}$ & 2.06 & 0.06 \\pt]
$B2 / {\rm kG}$ & 0.17 & 0.07 \\
$\Phi_2 / {\rm rad}$ & 1.6 & 0.4 \\
$\beta / ^\circ$ & 62 & $+5/-13$ \\
$B_{\rm d} / {\rm kG}$ & 4.0 & $+2.8/-0.2$ \\
\hline
$B_{\rm K} / {\rm kG}$ & 0.30 & $+0.10/-0.02$ \\
$\log{\dot{M} / {\rm M_\odot~yr^{-1}}}$ & $-9.7$ & 0.1 \\
$v_\infty / {\rm km~s^{-1}}$ & 1140 & 20 \\[2.5 pt]
$\log{\eta_*}$ & 6.0 & $+1.2/-0.1$ \\[2.5 pt]
$R_{\rm A} / R_*$ & 10 & $+5/-1$ \\
$\log{R_{\rm A}/R_{\rm K}}$ & 1.02 & $+0.07/-0.03$ \\
$\tau_{\rm J} / {\rm Myr}$ & 6 & 1 \\
$t_{\rm S,max} / {\rm Myr}$ & 5 & 1 \\
\hline\hline
\end{tabular}
\end{table}

\cite{2020MNRAS.493.4657L} determined the rotational period of $\rho$\,Oph\,A to be $P_{\rm rot} = 0.747326(2)$~d using the equivalent width curve of the He~{\sc i}~5015 line. In order to ensure an accurate phasing of the new magnetic data with previously acquired datasets, we revisit this rotational period in an attempt to attain a higher precision. \cite{2018AJ....155..196R} determined the rotational period of $\rho$\,Oph\,C from its {\em K2} light curve. As a first step, the {\em K2} light curve of $\rho$\,Oph\,C was acquired from MAST and analyzed using {\sc period04}. The periodogram is shown in the top panel of Fig.\ \ref{rhooph_kepler}. 

After pre-whitening with the rotational frequency of $\rho$\,Oph\,C and its first harmonic, which dominate the periodogram, it is apparent that there are a large number of frequencies unrelated to harmonics of $\rho$\,Oph\,C's rotational frequency. Closer investigation reveals that many of these frequencies are harmonics of $\rho$\,Oph\,A's rotational frequency, indicating that the light curve is contaminated by stray light from the much brighter $\rho$\,Oph\,A. Many of the others are unrelated to either star's rotational frequency. Given the crowded field, it is possible that these reflect rotational or pulsational modulation of other nearby stars, or systematics in the {\em K2} light curve itself; their amplitudes however are generally very low. 

The bottom panels of Fig.\ \ref{rhooph_kepler} show the light curves of $\rho$\,Oph\,A and C, folded with their respective rotational frequencies after being pre-whitened with frequencies unrelated to their respective rotational frequencies. Harmonic models are shown with each light curve. $\rho$\,Oph\,A's light curve shows what appears to be an eclipse, while that of $\rho$\,Oph\,C is characterized by a sharp brightening followed by a broad, flat light curve minimum. For $\rho$\,Oph\,A, this analysis gives a rotational period of 0.7482(4)~d, consistent within 2$\sigma$ with the 0.747326(2)~d period found by \cite{2020MNRAS.493.4657L} using the larger uncertainty from the {\em K2} analysis. Since the \cite{2020MNRAS.493.4657L} ephemeris evidently cannot be improved upon at this point, we adopt it without modification; as seen in the following section, it in any case provides a good phasing of the new and previously published magnetic data.

\section{Magnetometry}\label{sect:magfld}

   \begin{figure*}[hbt!]
   \centering
   \includegraphics[width=0.95\textwidth]{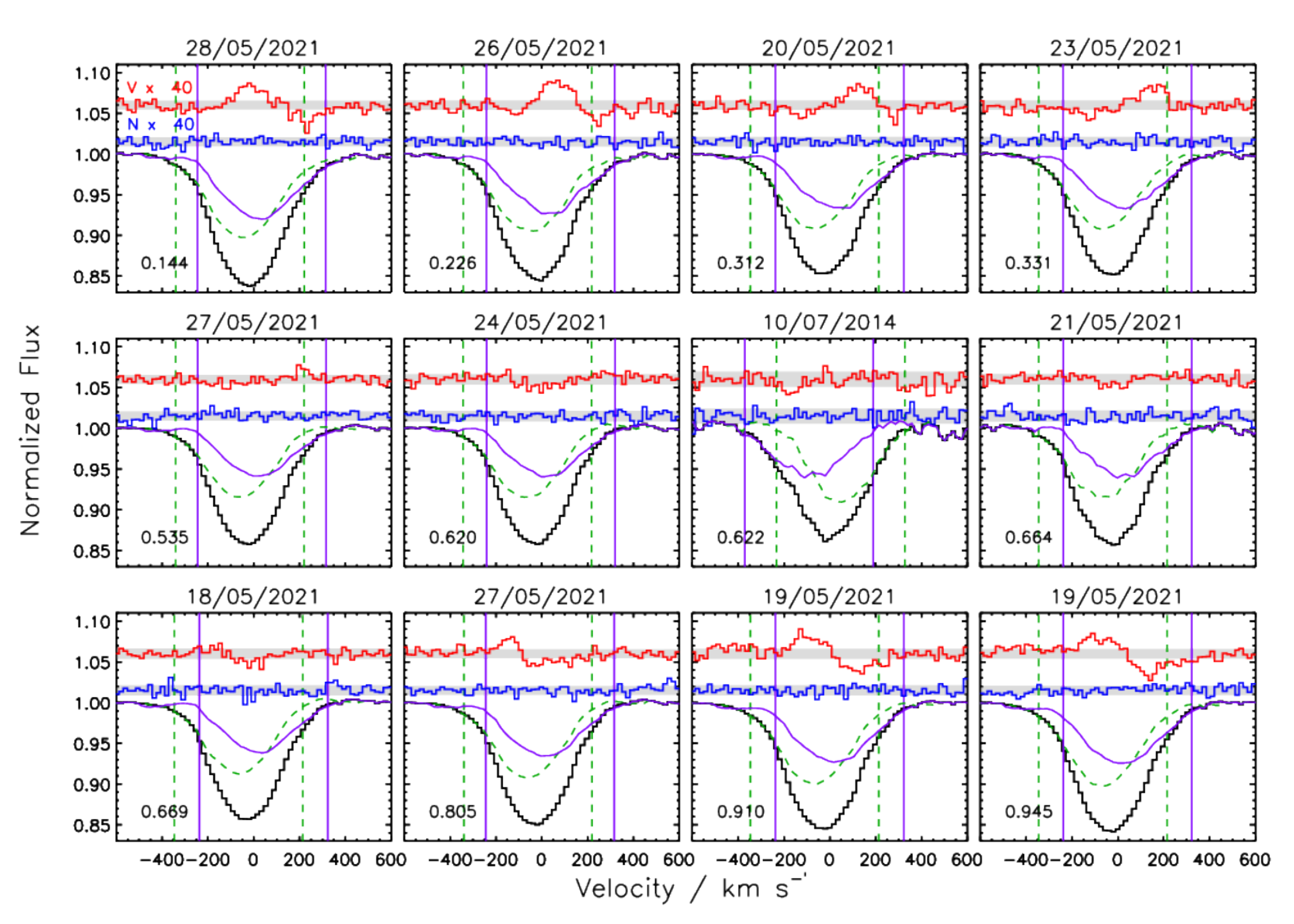}
      \caption[]{LSD Stokes $I$ (black), $N$ (blue), and Stokes $V$ (red) profiles of $\rho$\,Oph\,A. Each panel is labeled with the date of observation (top) and the rotational phase (lower left). The top left panel indicates the amplification applied to $N$ and Stokes $V$ for display purposes; these profiles have also been vertically offset for display. Dashed green and solid purple curves indicate the disentangled Stokes $I$ profiles of Aa and Ab. Vertical dashed green and solid purple lines show the integration ranges centered on the Aa and Ab radial velocities.}
         \label{rhoOphA_LSD}
   \end{figure*}

   \begin{figure}[hbt!]
   \centering
   \includegraphics[width=0.45\textwidth]{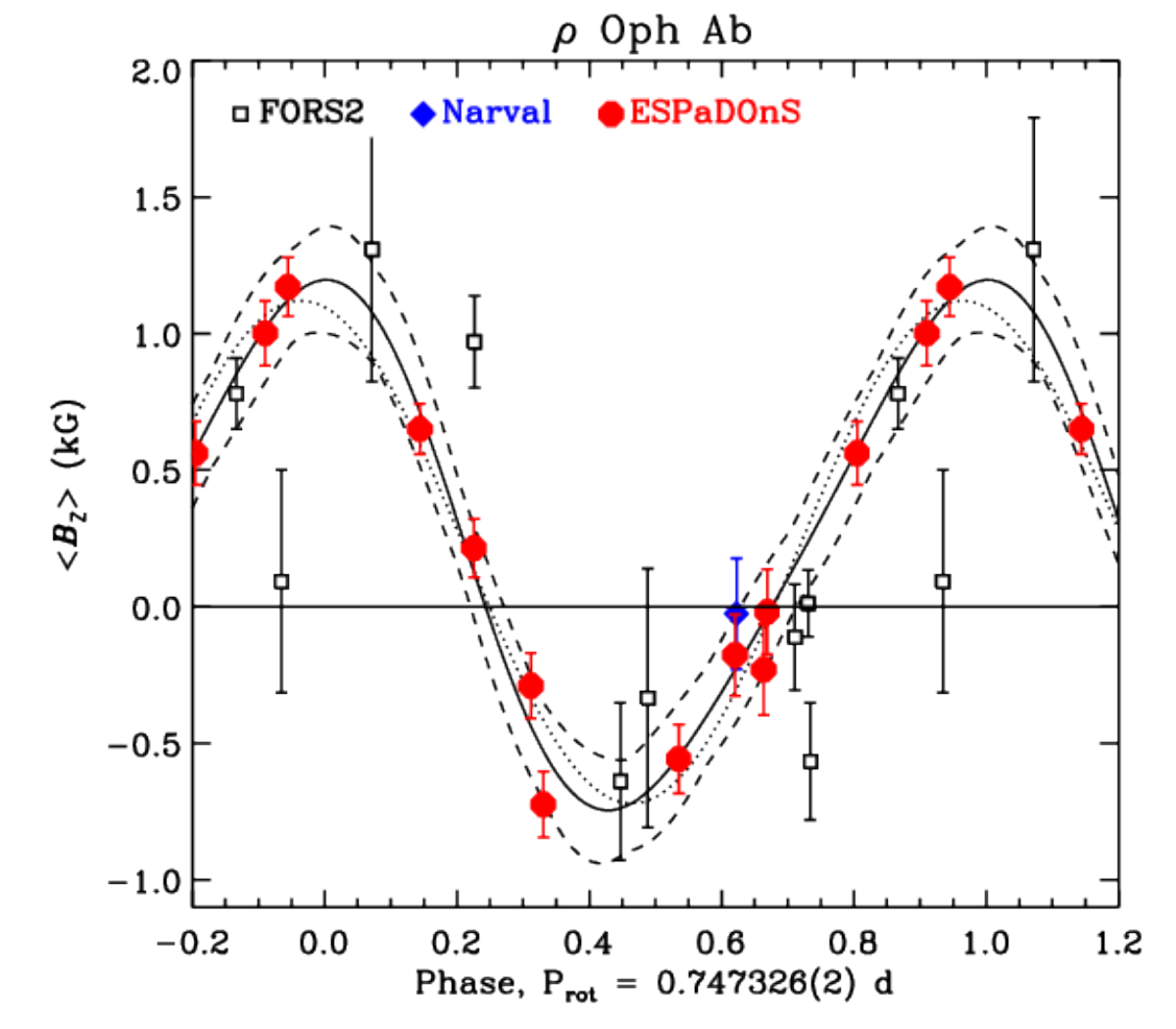}
      \caption[]{Longitudinal magnetic field \bz~of $\rho$\,Oph\,Ab measured from the LSD profiles shown in Fig.\ \ref{rhoOphA_LSD}, folded with the rotation period. Solid and dashed curves show the best harmonic fit and the fit uncertainty respectively. The dotted curve shows a sinsuoidal fit for comparison.}
         \label{rhoOph_bz}
   \end{figure}

   \begin{figure*}[hbt!]
   \centering
   \includegraphics[width=\textwidth]{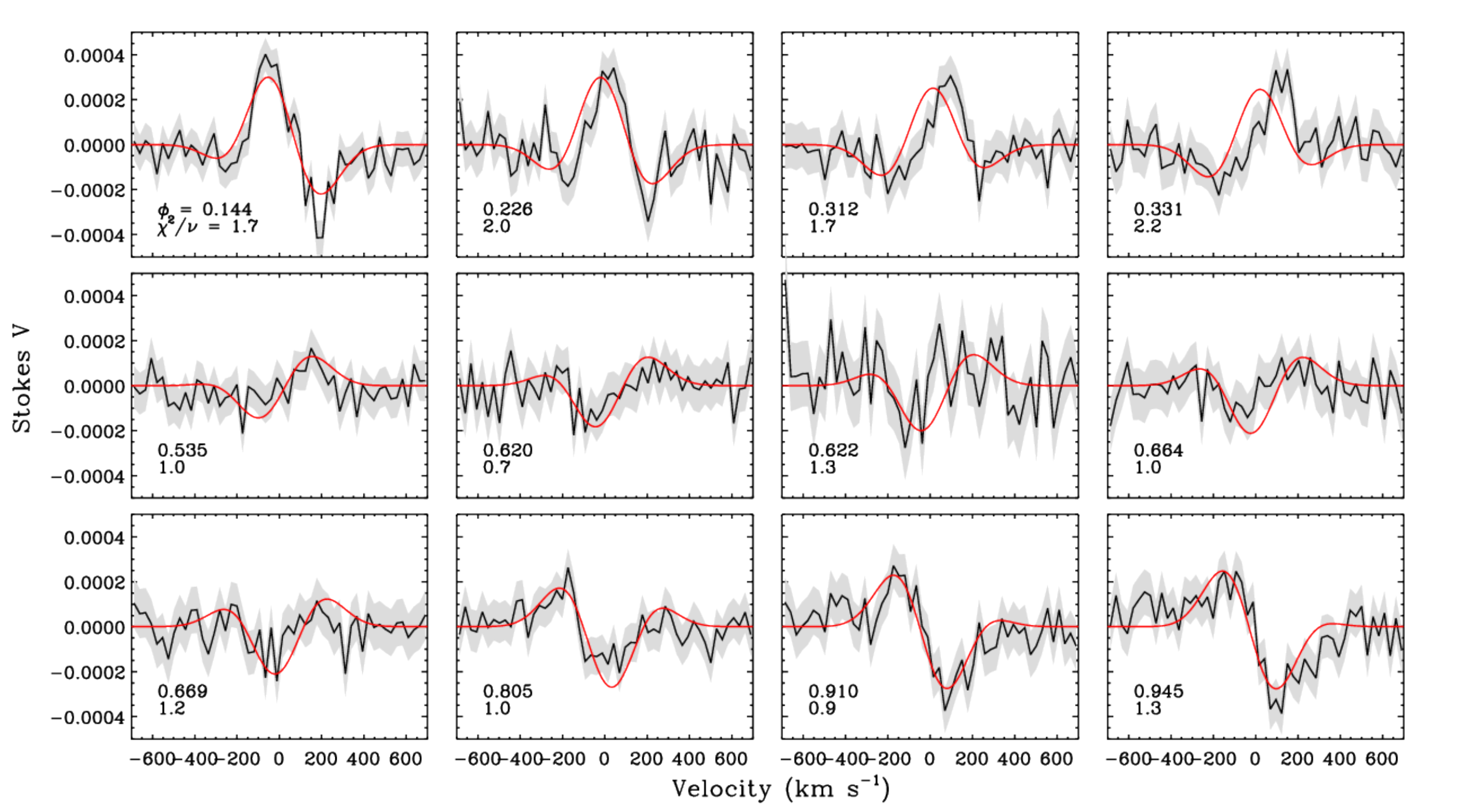}
      \caption[]{Comparison between observed LSD Stokes $V$ profiles (black, uncertainty in gray) and model Stokes $V$ inferred from the \bz~curve and the MCHRD sampler.}
         \label{rhoopha_lsd_modelv}
   \end{figure*}

In order to evaluate the magnetic field at the highest possible precision, mean line profiles were extracted from the spectra using least squares deconvolution \citep[LSD;][]{d1997}, in particular with the iLSD package developed by \cite{koch2010}. Line lists were obtained from the Vienna Atomic Line Database \citep[VALD3;][]{piskunov1995, ryabchikova1997, kupka1999, kupka2000,2015PhyS...90e4005R} via `extract stellar' requests with parameters matching those of the target stars. The line lists were then cleaned by eye in the usual way to remove H lines, lines strongly blended with the wings of H lines, and lines contaminated by telluric and interstellar lines, with the remaining lines then adjusted to match the depths of the lines in the observed spectrum \citep[the process is described in greater detail by][]{2018MNRAS.475.5144S}. 

The presence of a magnetic field was evaluated using false alarm probabilities (FAPs), and classified as definite, marginal, or non-detections (DD, MD, ND) according to the usual criteria of ${\rm FAP} < 10^{-5}$, $10^{-5} < {\rm FAP} < 10^{-3}$, and ${\rm FAP} > 10^{-3}$ \citep{d1997}, with FAPs measured using integration ranges set to the width of the Stokes $I$ spectral lines. The strength of the line-of-sight (longitudinal) magnetic field averaged over the visible stellar hemisphere \bz~was then measured using the centre-of-gravity method \citep[e.g.][]{mat1989}, with the same integration range as used for measuring FAPs. In addition to evaluation of the Stokes $V$ profiles, the same measurements were performed on the null $N$ profiles in order to check for spurious signals. The \bz~and \nz~measurements, and the corresponding detection flags from the FAPs, are given in Table \ref{table:bz}. 

Since $\rho$\,Oph\,A is an SB2, it is first necessary to determine to which star the magnetic field belongs. Two sets of LSD profiles were extracted, using solar metallicity line lists extracted with \teff~of 23 kK and 19 kK to reflect Aa and Ab respectively. After cleaning the line lists respectively contained 293 out of 502 total lines, and 237 out of 462 original lines. He lines were included in both masks due to the difficulty of detecting the magnetic field. LSD profiles were extracted using 27~\kms~velocity pixels, in order to further boost the $S/N$. The LSD profiles were then iteratively disentangled using the procedure described by \cite{2018MNRAS.475.5144S,2018MNRAS.475..839S,2019MNRAS.482.3950S}, with RVs fixed to the orbital model determined in Sect.~\ref{subsect:orbit}. 

The LSD profiles from the 19 kK mask are shown in Fig.\ \ref{rhoOphA_LSD}. Evaluation of FAPs yields DDs in 7/13 observations, with the remainder being NDs; all $N$ profiles give NDs. Concentrating on the DDs in which the Zeeman signature is clearly visible in Stokes $V$, it is apparent that it is offset from the centre of the total Stokes $I$ line profile. An integration range centred on the Aa component does not fully span the Stokes $V$ signature, while by constrast an integration range centred on Ab fully spans the Zeeman signature. This demonstrates that $\rho$\,Oph\,Ab, rather than Aa, is the magnetic star. Subsequent magnetic analysis was therefore carried out using the disentangled Stokes $I$ profiles and stellar parameters of $\rho$\,Oph\,Ab. 
 
Fig.\ \ref{rhoOph_bz} shows \bz~folded with the rotational period determined by \cite{2020MNRAS.493.4657L}, where we have used a zero point $T_0$ determined via a harmonic fit to the ESPaDOnS and Narval \bz~measurements, defined as the time of maximum \bz~one cycle before the first observation (see Table \ref{table:mag_pars}.) The harmonic function takes the form

\begin{equation}\label{eqn:bz}
\langle B_z\rangle(\phi) = \sum_{i=0}^{n} B_i \sin{(i\phi + \Phi_i)},
\end{equation}

\noindent where $\phi$ is the rotational phase, and $B_i$ and $\Phi_i$ are respectively the amplitude and phase offset of the $i^{\rm th}$ term, with phase terms in radians. Optimal fits were determined by calculating the reduced $\chi^2$ of the fit $\chi^2/\nu$ (where $\nu$ is the number of degrees of freedom): if $\chi^2/\nu$ is approximately 1, the fit is considered good; signficantly greater than 1, poor; and below 1, over-fit. Expansion of Eqn.\ \ref{eqn:bz} to $n=1$ is consistent with a dipolar magnetic field; if higher values of $n$ are required to achieve a reasonable fit, \bz~may be affected by higher-order contributions to the surface magnetic field, or the polarization signatures might be weighted by the variable surface brightness arising from chemical spots. 

Fitting with Eqn. \ref{eqn:bz} up to $n=1$ yields $\chi^2 / \nu = 2$, while using $n=2$ gives $\chi^2/\nu =  1.5$, favouring the higher-order fit. Extending to $n=3$ does not improve the fit quality, therefore we adopt the $n=2$ fit, the parameters of which are reported in Table \ref{table:mag_pars}. Both fits are shown in Fig.\ \ref{rhoOph_bz}. While the higher-order fit is favoured, the modifications to the \bz~curve from the $n=1$ fit are minor, indicating only slight contributions from the non-dipolar components of the surface field to \bz.

Also shown in Fig.\ \ref{rhoOph_bz} are the FORS2 \bz~measurements previously reported by \cite{2018A&A...610L...3P} and \cite{2020MNRAS.493.4657L}, where we have chosen the more precise measurements obtained from all metallic lines. These measurements and their uncertainties have been increased by a factor of 2, to account for the fact that they were originally made under the assumption that $\rho$\,Oph\,A is a single star, but are in fact formed in the photosphere of Ab and are therefore diluted by the contribution of Aa to the spectrum. This dilution is already accounted for in the high-resolution \bz~measurements by the use of disentangled Stokes $I$ profiles. As can be seen in Fig.\ \ref{rhoOph_bz}, there is a generally good agreement between the high- and low-resolution results, once this correction has been made.

The terms $B_0$ and $B_1$ from Eqn.\ \ref{eqn:bz} were used to solve the geometrical relationship developed by \cite{preston1967,preston1974} between the inclination angle of the rotational axis from the line of sight $i_{\rm rot}$ and the obliquity angle $\beta$ between the rotational and magnetic axes for a tilted dipole:

\begin{equation}\label{preston_r}
r = \frac{|B_0| - B_1}{|B_0| + B_1} = \frac{\cos{(i_{\rm rot} + \beta)}}{\cos{(i_{\rm rot} - \beta)}},
\end{equation}

\noindent with $i_{\rm rot}$ independently constrained via knowledge of the stellar radius $R_*$ and the rotational broadening $\sin{i_{\rm rot}}$~(see Table \ref{table:stellar_pars}):

\begin{equation}\label{eqn:irot}
\sin{i_{\rm rot}} = \frac{v\sin{i_{\rm rot}}}{v_{\rm eq}} = \frac{P_{\rm rot}v\sin{i_{\rm rot}}}{2\pi R_*},
\end{equation}

\noindent where $v_{\rm eq}$ is the equatorial rotational velocity. The surface strength of the magnetic dipole $B_{\rm d}$ was then found via \citep{preston1967,preston1974}:

\begin{equation}\label{eqn:bd}
B_{\rm d} = \langle B_z \rangle_{\rm max} \frac{20(3 - \epsilon)(\cos{\beta}\cos{i_{\rm rot}} + \sin{\beta}\sin{i_{\rm rot}})}{15 + \epsilon},
\end{equation}

\noindent where $\epsilon$ is the limb-darkening coefficient, obtained from the tables calculated by \cite{2016MNRAS.456.1294R}, using the $B$ band values as those closest to the wavelength region containing the majority of spectral lines used in the analysis. 

In practice $i_{\rm rot}$, $\beta$, and $B_{\rm d}$ were calculated simultaneously using the Monte Carlo Hertzsprung Russell Diagram (MCHRD) sampler described by \cite{2019MNRAS.490..274S}, which combines all available atmospheric, rotational, and magnetic data with evolutionary models to obtain fully self-consistent fundamental, rotational, magnetic, wind, and magnetospheric parameters satisfying physical constraints such as under-critical rotation and oblateness (expressed here as the ratio $R_{\rm pol}/R_{\rm eq}$ of the polar to equatorial radii). Asymmetric uncertainties are obtained directly from posterior distributions, and automatically account for correlations between variables. 

Rotational and magnetic parameters inferred from the MCHRD sampler are given in Table \ref{table:mag_pars}. The star is slightly oblate ($R_{\rm pol}/R_{\rm eq} = 0.94$), with a high inclination ($i_{\rm rot} = 73^\circ$) and moderate obliquity ($\beta = 62^\circ$). Its surface magnetic field $B_{\rm d} = 4$~kG, consistent with the lower range of magnetic field strengths typically seen in young, rapidly rotating magnetic B-type stars \citep{2019MNRAS.490..274S}. These are significant modifications to the properties determined by \cite{2020MNRAS.493.4657L}, who found $R_{\rm pol}/R_{\rm eq} = 0.86$, $i_{\rm rot} = 35^\circ$, $\beta = 80^\circ$, and $B_{\rm d} = 2.7$~kG. These differences are all explained as a consequence of the smaller mass and radius of $\rho$\,Oph\,Ab as compared to the values adopted by \cite{2020MNRAS.493.4657L}, as well as the stronger \bz~measurements obtained once the contribution of $\rho$\,Oph\,Aa is corrected for. 

To check the validity of the magnetic model inferred from the \bz~curve and the stellar and rotational parameters obtained from the MCHRD sampler, Fig.\ \ref{rhoopha_lsd_modelv} compares the observed Stokes $V$ profiles to synthetic profiles generated with the model's parameters. The synthetic Stokes $V$ profiles were calculated using $rho$\,Oph\,Ab's \vsini, with an additional Gaussian broadening applied to the local line profiles used for disk integration (necessary due to the inclusion of He lines in the line mask), normalized to the EW of the individual disentangled Stokes $I$ profiles, and shifted to $\rho$\,Oph\,Ab's RV. The reduced $\chi^2$ is close to 1 (1.07) averaged across all fits, with values for individual fits ranging between 0.7 and 2.2 (see labels in Fig.\ \ref{rhoopha_lsd_modelv}, indicating that the simple tilted dipole model derived above provides a good fit to the observations. 

\section{Magnetosphere}\label{sect:magnetospheres}

   \begin{figure*}[hbt!]
   \centering
   \includegraphics[width=\textwidth]{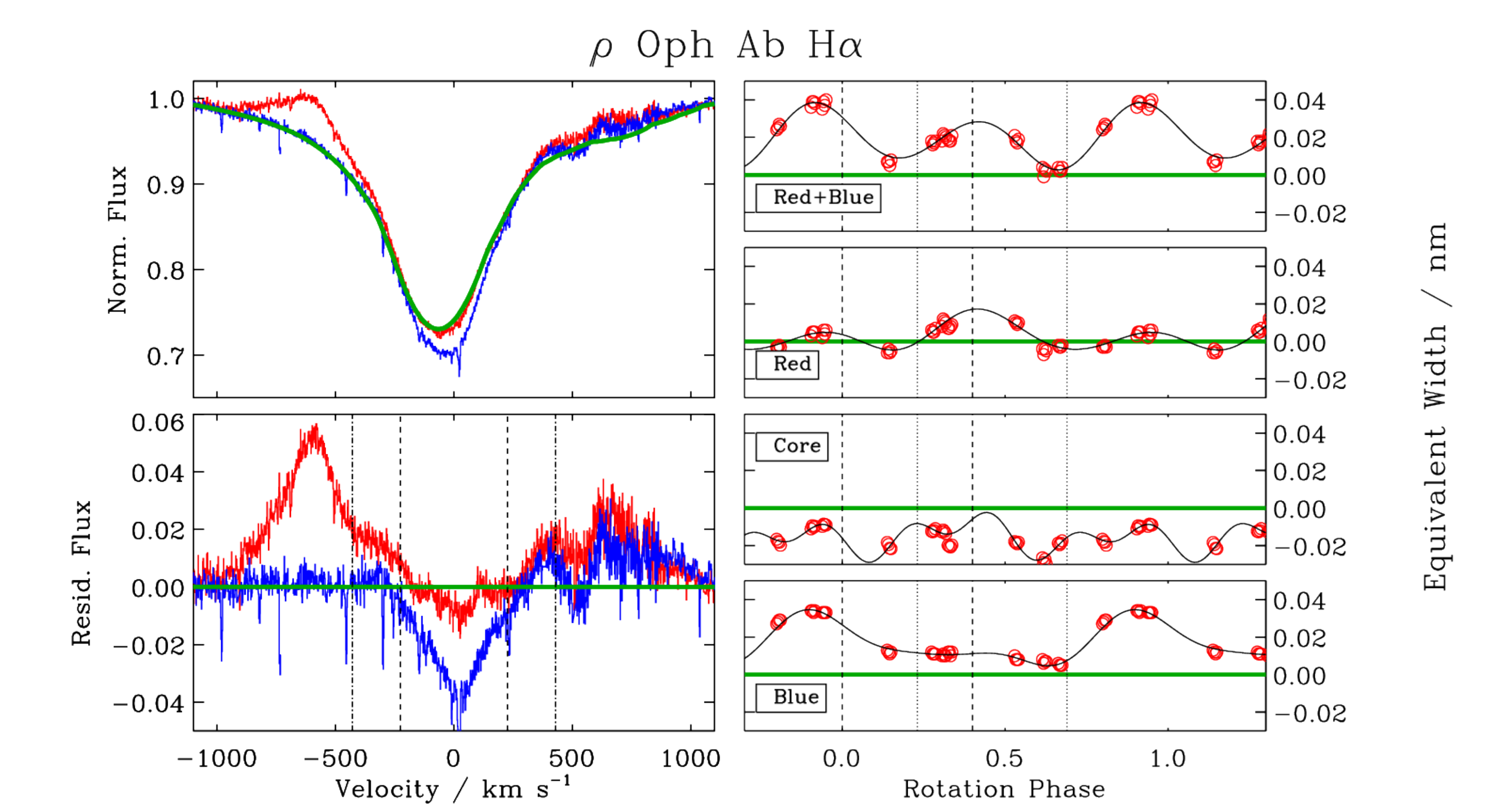}
      \caption[]{{\em Top left}: H$\alpha$ profiles of $\rho$\,Oph\,Ab at maximum emission (red) and minimum emission (blue), shown with a synthetic spectrum (thick green). {\em Bottom left}: residual flux after subtracting the synthetic spectrum from the observed spectra. Horizontal green line indicate the continuum level. Vertical dashed and dot-dashed lines show $\pm$\vsini~and the velocities corresponding to $\pm R_{\rm K}$, respectively (since the magnetosphere cororates with the star, there is a direct linear relationship between line-of-sight velocity and projected distance). {\em Right panels}: equivalent widths (EWs) measured from (bottom to top) the blue wing, the line core, the red wing, and the combined blue and red EWs. Dashed and dotted vertical lines show \bz~extrema and nulls from the fit in Fig.\ \ref{rhoOph_bz}. Horizontal thick green lines show the continuum level. Curves show harmomic fits.}
         \label{rhoOphA_halpha_multiplot}
   \end{figure*}

The magnetospheric parameters inferred from the MCHRD sampler are reported in Table \ref{table:mag_pars}. The magnetospheres of early-type stars are primarily characterized by two radii, the Kepler corotation radius \rk~and the Alfv\'en radius \ra~\citep[e.g.][]{petit2013,2019MNRAS.490..274S}. The Alfv\'en radius gives the maximum extent of magnetic confinement in the magnetic equatorial plane, and scales with the dimensionless wind magnetic confinement parameter $\eta_* \equiv B_{\rm eq}^2R_*^2\dot{M}^{-1}v_\infty^{-1}$, as $R_{\rm A} / R_* \sim \eta_*^{1/4}$, where $\eta_*$ expresses the ratio of magnetic to wind kinetic energy density at the surface in the vicinity of the magnetic equator, where the wind and magnetic field are perpendicular to one another \citep{ud2002,ud2008}. If $\eta_* > 1$ the wind is considered to be magnetically confined, i.e.\ the star has a magnetosphere. For $\rho$\,Oph\,Ab, $\eta_* > 10^6$, i.e.\ it is strongly magnetically confined (indeed, within the region of applicability of the RRM model), where the mass-loss rate $\dot{M}$ and wind terminal velocity $v_\infty$ were evaluated using the \cite{vink2001} recipe\footnote{If the \cite{krticka2014} mass-loss rates are used instead, $\eta_*$ is even larger since the \citeauthor{krticka2014} $\dot{M}$ is much lower.}.

Since magnetically trapped plasma corotates with the star, it is subjected to a centrifugal force that increases with distance from the axis of rotation. The Kepler radius is the distance from the rotational axis at which the centrifugal and gravitational forces are equivalent. As predicted by the semi-analytic Rigidly Rotating Magnetosphere \citep[RRM;][]{town2005c} model, and verified via magnetohydrodynamic (MHD) simulations \citep{ud2006,ud2008,ud2009}, material flowing into the magnetosphere that accumulates below \rk~(also given in units of $R_*$) will fall back to the surface of the star on dynamical timescales, while if the stellar rotational velocity and magnetic field strength are both sufficiently high that $R_{\rm K} < R_{\rm A}$, material between the two radii will accumulate to high densities before being expelled outwards by centrifugal breakout \citep{ud2006,ud2008,2020MNRAS.499.5366O}. This region is referred to as the centrifugal magnetosphere; if it is sufficiently large and dense, H$\alpha$ emission can occur even in B-type stars, whose winds are generally too weak to result in detectable emission \citep[e.g.][]{petit2013,2020MNRAS.499.5379S}.

$\rho$\,Oph\,Ab has $R_{\rm A} > R_{\rm K}$, and indeed $\log{R_{\rm A}/R_{\rm K}} > 1$, i.e.\ it is well within the regime in which H$\alpha$ emission is expected for magnetic B-type stars \citep[e.g.][]{petit2013,2019MNRAS.490..274S}. A second numerical indication that H$\alpha$ emission should be present is provided by the strength $B_{\rm K} = B_{\rm eq}/R_{\rm K}^3$ of the equatorial magnetic field at $R_{\rm K}$. \cite{2020MNRAS.499.5379S} showed that all magnetic B-type stars with $B_{\rm K} > 100$~G display H$\alpha$ emission. $\rho$\,Oph\,Ab is well above this threshold (see Table \ref{table:mag_pars}).

\subsection{H$\alpha$ magnetosphere}\label{subsect:halpha}

The radio and X-ray magnetosphere of $\rho$\,Oph\,A has already  been studied in detail \citep{2016A&A...592A..88P,2017A&A...602A..92P,2020MNRAS.493.4657L}. However, H$\alpha$ emission has not previously been reported. Our examination of the previously reported spectroscopic dataset of this star revealed that the star's H$\alpha$ emission has likely gone unremarked due to the low quality of the available spectra in the vicinity of H$\alpha$, which suffer from poor $S/N$, distortions introduced by poor continuum normalization, over-exposure, etc.; we therefore base our analysis entirely on the new ESPaDOnS dataset. 

H$\alpha$ was evaluated using the method described by \cite{2020MNRAS.499.5379S}. Synthetic spectra (created in the same fashion as used to measure atmospheric parameters in Sect. \ref{subsec:atmos}, i.e.\ from TLUSTY spectra shifted to the RVs of the two components and added with a fixed luminosity ratio) were subtracted from the observed profiles, with the equivalent width measured from the furthest blue extent of emission to $-$\vsini, between $\pm$\vsini, and from $+$\vsini~to the furthest red extent of emission. This method enables measurement of the equivalent width of the emission component, unaffected by absorption from plasma projected in front of the star, and evaluation of the relative contributions of different parts of the magnetosphere to the total emission. 

The H$\alpha$ emission profiles at minimum and maximum emission are shown together with the synthetic spectrum in Fig.\ \ref{rhoOphA_halpha_multiplot}. $\rho$\,Oph\,Ab has a strong emission bump in the blue wing of the profile, with the peak emission occurring at about $-600$~\kms, corresponding to about 2.6~$R_*$ given the \vsini~of the magnetic component and using the assumed corotation of the circumstellar plasma to convert Doppler velocity to projected distance. There is no obvious emission below \rk~at 1.9~$R_*$. This is consistent with the general pattern expected from CMs, and is a direct prediction of the circumstellar plasma distribution expected from the RRM model. 

At minimum emission, there is enhanced absorption in the line core of H$\alpha$, which may be indicative of eclipsing of the star by the CM. This is another prediction of the RRM model for a star with $\rho$\,Oph\,Ab's large $i+\beta > 90^\circ$. The double-wave character of the EW variation across the line (right panels of Fig.\ \ref{rhoOphA_halpha_multiplot}) is also consistent with expectations from the RRM model. The two emission maxima closely coincide with the positive and negative extrema of the \bz~curve, with the peak emission strength coinciding with the stronger positive \bz~extremum at phase 0. Emission minima correspond to \bz~nulls, at which phases there is enhanced absorption in the line core. In the RRM framework, this is explained by the plasma trapped approximately in the magnetic equator, being seen most nearly face-on when the magnetic pole is closest to the centre of the visible hemisphere of the star (i.e.\ \bz~maximum), and being seen again at a somewhat greater tilt relative to the line of sight when the second magnetic pole comes closest to the centre of the visible hemisphere. On the other hand, emission minima occur when the magnetic equator bisects the visible hemisphere, i.e.\ a \bz~null, at which phase the magnetosphere is projected in front of the star, thereby leading to an eclipse. 

Examining the blue and red EW curves individually, it is clear that the EW variation is dominated by one cloud, which is seen most nearly face-on near phase 0.9 in the blue wing, and then shows a second maximum one half of a rotational cycle later in the red wing. This is also apparent from the H$\alpha$ profiles in Fig.\ \ref{rhoOphA_halpha_multiplot}. Notably, emission maxima and minima do not correspond precisely to \bz~extrema and nulls, but precede them by about 0.1 cycles; similar behaviour is seen in the H$\alpha$ EW curves of other CM stars \citep{2020MNRAS.499.5379S}.

\subsection{Light curve}\label{subsubsect:rhoopha_lightcurve}

   \begin{figure}
   \centering
   \includegraphics[width=0.45\textwidth]{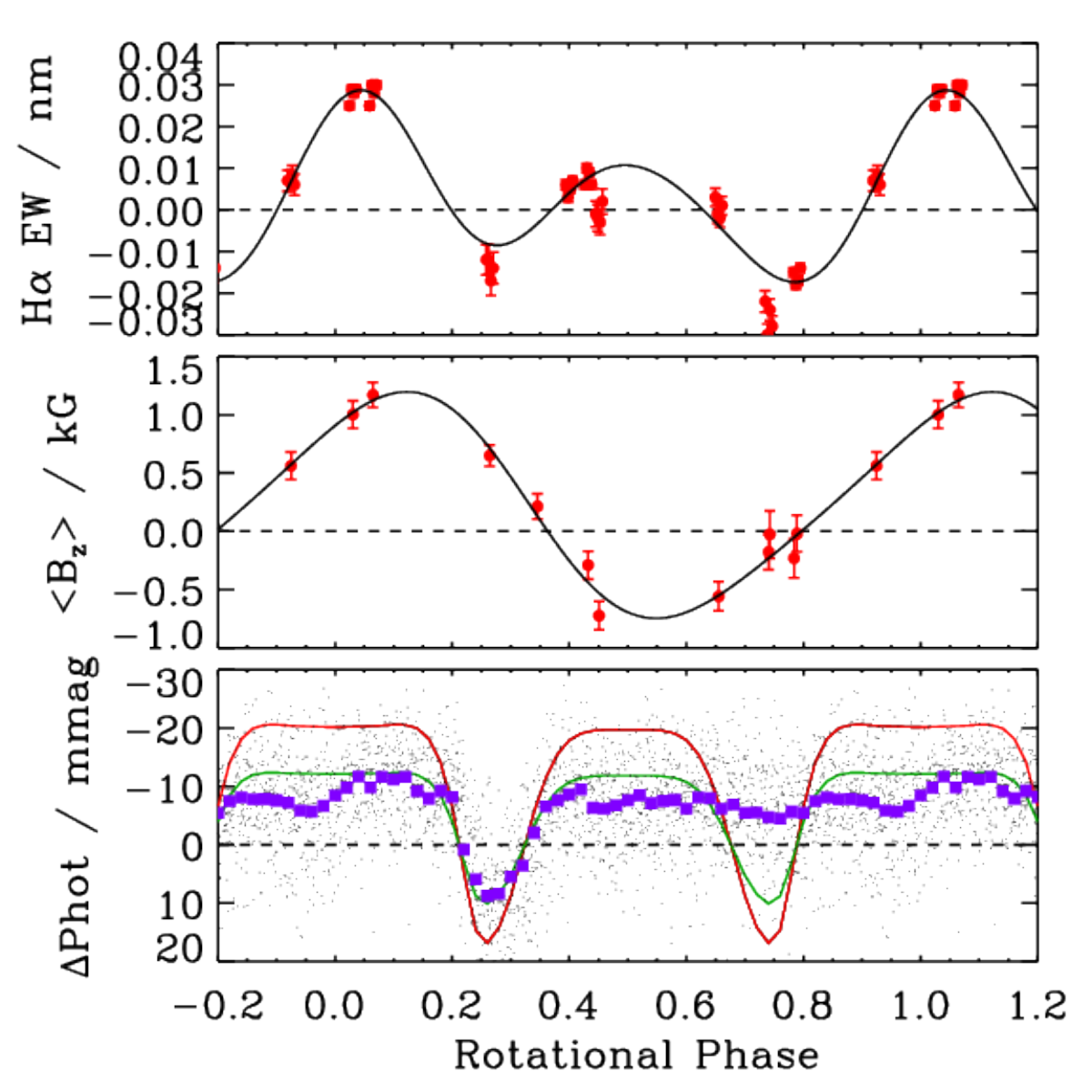}
      \caption[]{{\em Top}: H$\alpha$ emission EWs across the full line (i.e.\ core plus wings) for $\rho$\,Oph\,Ab folded with the rotational phase. The dashed line indicates the continuum. The solid curve shows a harmonic fit. {\em Middle}: \bz~measurements. The dashed line shows the magnetic null. The solid curve shows a harmonic fit. {\em Bottom}: {\em K2} light curve adjusted to remove dilution from Aa, B, and C. Phase-binned data are shown by purple squares. Red and green curves show CBO-RRM models with $\tau_{\rm K} = 0.1$ and 0.06 respectively.}
         \label{rhoOphA_K2}
   \end{figure}

Rotationally modulated photometric variability is essentially ubiquitous in early-type magnetic stars. In most cases it arises due to surface chemical abundance spots, however amongst the early B- and O-type stars, photometric variation can also arise due to the magnetosphere, primarily due to eclipses of the star by dense magnetospheric clouds. The best-known example of this phenomenon is $\sigma$\,Ori\,E \citep{lb1978,town2005b}. Eclipse models have also reproduced the light curves of magnetic O-type stars \citep{2020MNRAS.492.1199M}. \cite{2022A&A...659A..37K} showed that complex photometric variability, characterized by a large number of sharp features, can be explained as eclipses originating in a magnetosphere formed in a complex magnetic field superimposed on the smoother variability originating with surface spots. 

Fig.\ \ref{rhoOphA_K2} shows the H$\alpha$ emission EW, \bz, and light curve of $\rho$\,Oph\,Ab, folded on the rotation period determined by \cite{2020MNRAS.493.4657L}. In order to determine as accurately as possible $\rho$\,Oph\,Ab's true photometric amplitude, we re-extracted the {\em K2} light curve, with apertures centred on $\rho$\,Oph\,C and $\rho$\,Oph\,A. The resulting light curve of $\rho$\,Oph\,C provided a good match to the {\em Hipparcos} light curve, suggesting that this attempt to separate the light curves was mostly successful. Fourier analysis however revealed residual contamination of the $\rho$\,Oph\,A light curve with a contribution from $\rho$\,Oph\,C of about 5\% of the amplitude of the $\rho$\,Oph\,C light curve, indicating that about 5\% of the light in the $\rho$\,Oph\,A light curve comes from the other star. Due to the {\em Kepler} pixel size, the contributions of $\rho$\,Oph\,B could not be isolated. Instead, its contribution was estimated from the $V$ magnitudes of the A and C components. Finally, the relative contributions of Aa and Ab were determined from evolutionary models and the system's orbital properties (see Sect.\ \ref{subsubsect:rhoopha_fundpars}). In the end, we estimate $\rho$\,Oph\,Ab to contribute about 20\% of the light in the {\em K2} light curve, and normalized accordingly. 

As can be seen in Fig.\ \ref{rhoOphA_K2}, the light curve is characterized by a sharp dip centred on phase 0.25. Outside of this, the variability is of a much lower amplitude, with some suggestion of a second dip at phase 0.75. These dips correspond to approximately the same phases as nulls in \bz~and minima in H$\alpha$; as can be seen in Fig.\ \ref{rhoOphA_halpha_multiplot}, the H$\alpha$ minima exhibit eclipses in the rotationally broadened core of the line. The asymmetry in H$\alpha$, with one of the emission bumps being substantially stronger than the other, is also consistent with the asymmetry in eclipse depth. Further, the deepest eclipse corresponds to the phase at which the strongest emission bump crosses from blue to line centre, i.e.\ when the largest cloud moves in front of the star. The shallower photometric eclipse corresponds to the weaker emission bump moving from blue to line centre, i.e.\ eclipsing by a smaller cloud. The photometric and spectroscopic behaviour of $\rho$\,Oph\,Ab is therefore very similar to that exhibited by $\sigma$\,Ori\,E, making $\rho$\,Oph\,Ab one of the very few stars exhibiting $\sigma$\,Ori\,E-type photometric variability. 

To investigate this further, we attempted to model the light curve using the Centrifugal Breakout--Rigidly Rotating Magnetosphere (CBO-RRM) model developed by \cite{2022MNRAS.511.4815B}. This model extends the original photometric RRM model \citep[e.g.][]{town2005c,town2008} to account for the density scaling inferred from CBO \citep{2020MNRAS.499.5366O}, and to include photometric emission via Thomson scattering. The latter was motivated by the out-of-eclipse brightening exhibited by $\sigma$\,Ori\,E, which \cite{2015MNRAS.451.2015O} showed could not be explained via photometric spots. Photometric emission was also included by \cite{2020MNRAS.492.1199M}, although they used the original $r^{-3}$ RRM density scaling rather than the $r^{-5}$ CBO scaling, and approximated photometric emission as a reflection effect rather than solving the radiative transfer equations as in the models presented by \cite{2022MNRAS.511.4815B}. 

To synthesize the model, we adopted $i$, $\beta$, and the dimensionless critical rotation parameter $W$ \citep[defined as the ratio of the surface equatorial rotational velocity to the velocity required to maintain a Keplerian orbit at the surface, see][]{ud2008} as inferred from the modelling described in Sect.\ \ref{sect:magfld} (Table \ref{table:mag_pars}). Phase 0 in the models corresponds to magnetic maximum, i.e.\ the same definition as used for $T0$ in Table \ref{table:mag_pars}. The optical depth at the Kepler radius $\tau_{\rm K}$ was initially determined from Eqn.\ 7, \cite{2022MNRAS.511.4815B}:

\begin{equation}\label{tauk}
\tau_{\rm K} = 0.11 \frac{c_{\rm f}}{0.3}\left (\frac{B_{\rm d}}{\rm kG} \right )^2 W^{8/3},
\end{equation}

\noindent where $c_{\rm f}$ is a `correction factor' inferred by \cite{2020MNRAS.499.5366O} from magnetohydrodynamic models and taken to be about 0.3. With uncertainties in $W$ and $B_{\rm d}$, and assuming the value of $c_{\rm f}$ adopted by \cite{2022MNRAS.511.4815B} we find that $\tau_{\rm K} = 0.12^{+0.32}_{-0.02}$. The red curve in the bottom panel of Fig.\ \ref{rhoOphA_K2} shows a model with $\tau_{\rm K} = 0.1$, which is about twice the amplitude of the observed curve. Reducing this to $\tau_{\rm K} = 0.06$ (green curve) yields an excellent agreement with the depth and shape of the primary eclipse at phase 0.25, with the phase (albeit not the depth) reproduced at phase 0.75. However, we note that achieving this fit required an arbitrary phase shift of 0.12 to be applied to the {\em K2} light curve (with the same phase shift applied to the EWs and \bz~in Fig.\ \ref{rhoOphA_K2}). We also shifted the {\em K2} light curve by $-6$~mmag, a justified change since the continuum level is ambiguous in differential photometry. 

The factor of $\sim$2 discrepancy between the calculated and best-fit values of $\tau_{\rm K}$ is consistent with the results of the light curve fitting test performed by  \cite{2023MNRAS.523.6371B}, who found large discrepancies between the values of $\tau_{\rm K}$ required to reproduce the light curves of CM host stars, and the values calculated from Eqn.\ \ref{tauk}. However, it is notable that all 5 stars examined by \cite{2023MNRAS.523.6371B} are both predicted and measured to have much higher values of $\tau_{\rm K}$ (typically around unity) than those found here for $\rho$\,Oph\,Ab, i.e.\ $\rho$\,Oph\,Ab is predicted to have, and in fact does have, the shallowest magnetospheric eclipse of any star for which this phenomenon has been observed. 

\section{Discussion}\label{sect:discuss}

\subsection{$\rho$\,Oph\,A in the context of magnetic binaries}

With its identification as an SB2, $\rho$\,Oph\,A joins the short list of binary systems containing a magnetic star. The Binarity and Magnetic Interactions in various classes of Stars \citep[BinaMIcS;][]{2015IAUS..307..330A} survey demonstrated, via observations of 200 hot, close ($P_{\rm orb} < 30$~d) binaries, that fewer than 2\% of such systems contain a magnetic star. This remarkable result stands in stark contrast to the overall extremely high binary fraction for early-type stars. The properties of the known close magnetic binaries were summarized by \cite{2017A&A...601A.129L}. Since then, a few additional systems have been reported, e.g.\ the extremely tight, tidally-locked H$\alpha$-bright system HD\,156324 \citep{2018MNRAS.475..839S}, and the eclipsing binaries HD\,66051 and HD\,62658 \citep{2018MNRAS.478.1749K,2019MNRAS.490.4154S}. 

Since $\rho$\,Oph\,A apparently has an orbital period of at least a few months, it does not quite qualify as a close binary according to the BinaMIcS criterion. It is more analogous to the apparently pre-main sequence system W\,601 \citep{2021MNRAS.504.3203S}, which also consists of 2 early B-type stars with a somewhat eccentric orbit of a few months duration, and in which the magnetic field and magnetospheric H$\alpha$ emission also originate in the rapidly rotating, strongly magnetized secondary star. NU Ori is also somewhat similar: in this case, the system is a hierarchical triple, with an inner pair consisting of a non-magnetic B0 star and a non-magnetic late B-type star, and an outer, rapidly rotating, strongly magnetic B2 star with an orbit on the order of 1.5 years \citep{2019MNRAS.482.3950S}. In all 3 cases, the magnetic star is the less massive companion of a larger, non-magnetic central star. There are numerous counter-examples, however, in which it is the more massive primary that hosts the magnetic field, e.g. HD\,37017, $\delta$\,Ori\,C, or HD\,149277 \citep{1998AA...337..183B,leone2010,2018MNRAS.481L..30G}. 

What all of these systems have in common is that they challenge the notion that fossil magnetism is generally a relic leftover from dynamos powered by binary mergers \citep[e.g.][]{2019Natur.574..211S}, which has been invoked to reconcile the extremely low incidence of magnetic binaries, and the similar 10\% incidence rates of fossil magnetic fields and binary mergers. Producing a magnetic binary system via merger necessitates the system beginning its evolution as a hierarchical triple, with the inner binary merging due to some combination of stellar evolution and hardening of the orbit via dynamical interactions. While N-body simulations have demonstrated that the evolution of hierarchical triples is indeed expected to increase the merger rate substantially over the rate expected for binary systems \citep{2023A&A...678A..60K}, when the orbit of the outer star is not disrupted the resulting binary system is expected to be wide (with a minimum semi-major axis of 1 AU, and a median of around 46 AU) and highly eccentric (a median of around $e \sim 0.7$), reflecting the orbit of the outer companion \citep{2025A&A...693A..14B}. Indeed, the magnetic Of?p star HD\,148937 appears to be just such a system, and exhibits other signs of a recent merger event having produced the magnetic star \citep{2024Sci...384..214F}. To the contrary, there are systems with extremely tight orbits \citep[e.g.\ the tidally locked HD\,156324, with a circular orbit of around 1.5~d][]{2018MNRAS.475..839S}. The doubly magnetic system $\epsilon$\,Lupi, with two magnetic B-type stars in a 4.5~d orbit \citep{2015MNRAS.454L...1S}, is even more difficult to explain by a merger scenario, as this would require two mergers within a quadruple system. The orbital properties of $\rho$\,Oph\,A are just barely consistent with the end-products of triple evolution, corresponding to around 1\% of post-merger binaries.

\subsection{Spin-orbit alignment}

Since their rotational periods can be precisely determined, the rotational axis inclination angles of magnetic stars can also be tightly constrainted. For $\rho$\,Oph\,Ab we find $i_{\rm rot} = 73^{\circ+4}_{-6}$. This is consistent within uncertainty to the orbital inclination angle $i_{\rm orb} = 69 \pm 11^\circ$ inferred from comparison of the mass function to the masses inferred from atmospheric parameters and evolutionary models. Alignment of spin and orbital axes is expected for close binaries, as stars born from the same molecular cloud core should inherit parallel angular momentum vectors, which should express itself both in the orbit and the rotation of each star. However, spin-orbit alignment would be somewhat surprising in the end-product of a presumably chaotic triple-system merger event, casting further doubt on this scenario.

\subsection{Rotational evolution}

Magnetic stars are expected to rapidly shed angular momentum via their corotating magnetospheres \citep{ud2009}. Consistent with this, magnetic stars are generally much more slowly rotating than non-magnetic stars of comparable masses and evolutionary stages \citep{2018MNRAS.475.5144S}. 

The spindown age $t_{\rm S, max}$ is defined as the time necessary for a star to reach its present rotational velocity assuming initially critical rotation on the zero-age main sequence \citep[see Eqn.\ 13,][]{petit2013}. \cite{2020MNRAS.493.4657L} calculated $t_{\rm S,max} = 25 \pm 15$~Myr, which was difficult to reconcile with the estimated 5--10 Myr age of the $\rho$\,Oph group. With the corrected magnetic parameters determined here, this discrepancy is resolved: the stronger magnetic field gives $t_{\rm S, max} = 5 \pm 1$~Myr, perfectly consistent with  the age of the $\rho$\,Oph cluster. However, as the spindown timescale is highly sensitive to the mass-loss rate, adopting the lower \cite{krticka2014} \mdot~yields $t_{\rm S, max} = 21 \pm 5$~Myr, which is much older than the age of the cluster. There is also some remaining tension with the age of the stars themselves estimated from the HR diagram ($15 \pm 6$~Myr), which may point to shortcomings in non-magnetic evolutionary models \citep[e.g.][]{2022MNRAS.517.2028K}. 

Assuming that the rotational axis of the non-magnetic primary is also aligned with the orbital axis, as that of the magnetic secondary apparently is, it's interesting to note that the magnetic star seems to be more rapidly rotating than the primary. Since magnetospheric angular momentum loss should spin down the magnetic star more rapidly, this could indicate that the merger that produced the magnetic star may have happened relatively recently, within the last few Myr. Alternatively, since the stars are in any case young, $\rho$\,Oph\,Ab may simply have started its evolution as a more rapid rotator. 

\subsection{Magnetospheric asymmetry}

   \begin{figure}
   \centering
   \includegraphics[width=0.5\textwidth]{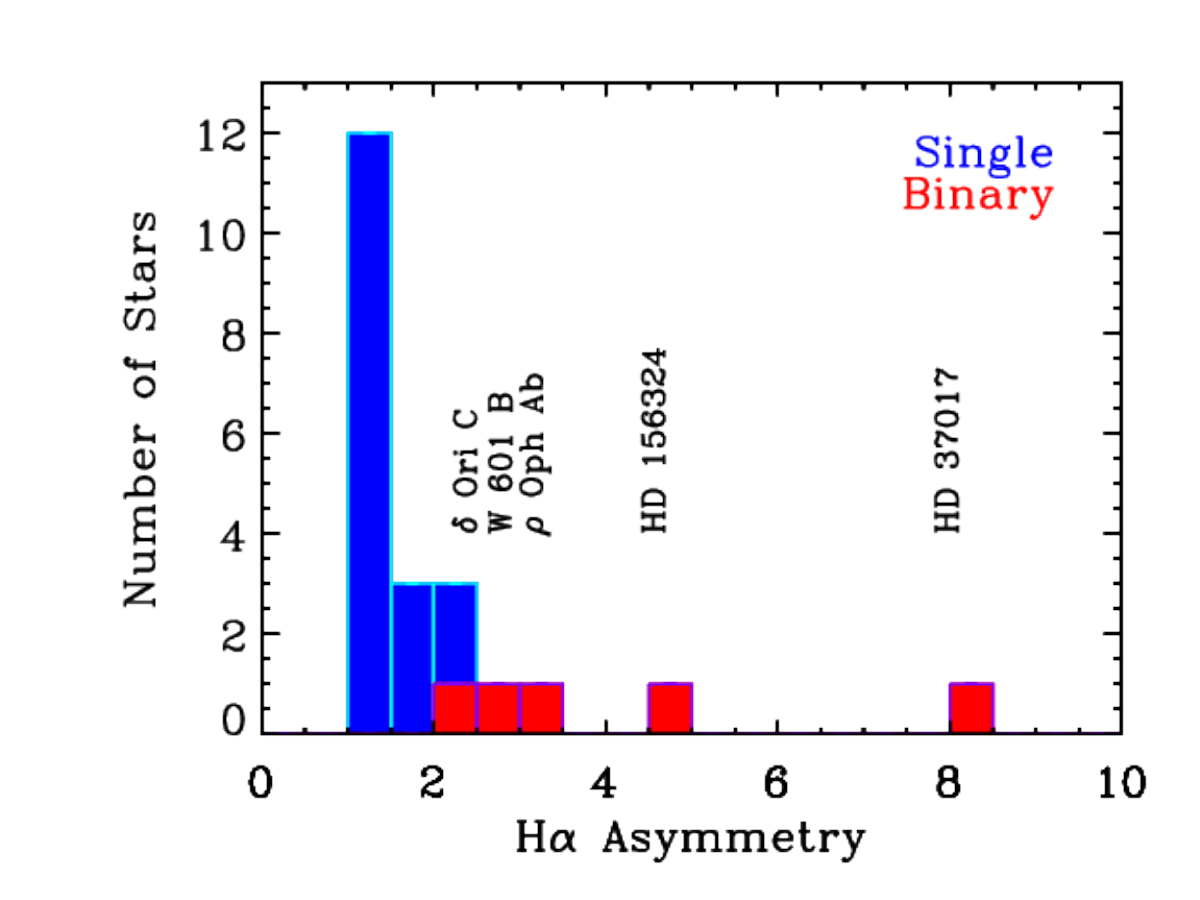}
      \caption[]{Asymmetry in the H$\alpha$ emission profiles of single(blue) and binary (red) magnetic B-type stars as measured from the ratio of blue to red equivalent widths at maximum emission (or red to blue, if red is stronger). A value of unity indicates a perfectly symmetric emission profile,}
         \label{halpha_vr_hist}
   \end{figure}

A tilted dipole should lead to a symmetrical magnetosphere, with two corotating clouds of equivalent size on either side of the magnetic axis, concentrated at the intersections of the magnetic and equatorial planes in the vicinity of the Kepler radius \citep{town2005c}. Indeed, this approximate configuration is commonly seen in the case of single magnetic stars \citep{2020MNRAS.499.5379S}. 

Some degree of asymmetry, with one cloud being larger than the other (and therefore producing more emission) is not uncommon -- indeed the prototypical centrifugal magnetosphere host star $\sigma$\,Ori\,E shows precisely this pattern \citep{oks2012}. Such asymmetry is usually explained by a magnetic field with significant departures from a pure dipole, with the strong quadrupolar component of $\sigma$\,Ori\,E's magnetic field being again an example of this \citep{2015MNRAS.451.2015O}. \cite{2020MNRAS.499.5379S} compared the $V/R$ ratios (i.e.\ the ratio of blue to red equivalent widths) of the H$\alpha$ emission profiles of centrifugal magnetosphere host stars to a simple empirical indicator of surface magnetic field complexity developed by \cite{2018MNRAS.475.5144S}, and found that stars with large departures of $V/R$ from unity (i.e.\ with highly asymmetric H$\alpha$ emission) also tended to have more complex magnetic fields, as would be predicted by the \cite{town2005c} RRM model. The exceptions to this were close binaries, which had simple magnetic fields but extreme emission profile asymmetry.

While there is some indication in \bz~that $\rho$\,Oph\,Ab's magnetic field is not a perfect dipole, the differences from a pure dipole are subtle, and may well be an artifact of the disentangling process used to obtain $\rho$\,Oph\,Ab's LSD Stokes $I$ profiles. Furthermore, the LSD Stokes $V$ profiles are apparently perfectly consistent with the dipolar model derived from \bz. Nevertheless, the star's H$\alpha$ emission profile is extremely asymmetrical, dominated by one large cloud, with the other almost totally absent (Fig. \ref{rhoOphA_halpha_multiplot}). This structure is also reflected in the photometric eclipses (Fig.\ \ref{rhoOphA_K2}). 

Fig.\ \ref{halpha_vr_hist} shows a histogram of H$\alpha$ profile asymmetry across the population of centrifugal magnetosphere host stars, as determined by the ratio of blue to red (or red to blue, if red is stronger) equivalent widths as measured at the phase of maximum emission \citep[where maximum emission corresponds to the phase at which the magnetosphere is seen closest to face-on, and emission profile asymmetry should be at a minimum;][]{town2005c}. For single stars, 2/3 of the population have essentially symmetric emission profiles, with the remainder showing relatively small degrees of asymmetry (one emission wing up to about twice as strong as the other, as in the case of $\sigma$\,Ori\,E).

As can be seen in Fig.\ \ref{halpha_vr_hist}, the H$\alpha$ emission profiles of magnetic binaries are systematically much more asymmetric than those of single stars. HD\,156324 is tidally locked, and the asymmetrical nature of its magnetospheric plasma distribution seems to be plausibly explained due to preferential accumulation of plasma around the Lagrange L3 point \citep{2018MNRAS.475..839S}. Neither HD\,37017 nor $\rho$\,Oph\,Ab are tidally locked \citep{1998AA...337..183B,2018MNRAS.475.5144S}, which complicates modeling, nor do they show evidence of any obvious harmonic relationship between the orbital and rotational periods. Their companions are also at much greater separations than that of HD\,156324, making a significant gravitational influence of the non-magnetic star on the magnetosphere somewhat less plausible. Nevertheless, the discovery of a highly asymmetric magnetosphere in a third binary system begins to make this phenomenon look less like a coincidence, and possibly deserving of theoretical attention. 

\section{Summary}\label{sect:summary}

$\rho$\,Oph\,A turns out to be a spectroscopic binary composed of two rapidly rotating early B-type stars with a slightly eccentric orbit of approximately 88 days. The previously reported magnetic field belongs to the secondary component, which we designate Ab. Since the previously determined stellar and magnetic parameters of this object were based on the assumption that it is a single star, these have been substantially revised. We find that $\rho$\,Oph\,Ab's 4 kG magnetic field is about twice as strong as the previously reported 2 kG, and the rotational axis is more highly inclined from the line of sight, approximately consistent with the inferred orbital inclination axis. The dipolar model inferred from \bz~and the stellar parameters provides an excellent match to the LSD Stokes $V$ profiles, suggesting that the surface magnetic field is quite close to a dipole.

We report H$\alpha$ emission for $\rho$\,Oph\,Ab for the first time. This emission was expected on the basis of the star's large centrifugal magnetosphere, but was not previously detected due to the subtle nature of the emission signature and the relatively low quality of previously available spectra in the vicinity of H$\alpha$. The emission is highly asymmetrical. The close match between a dipolar model of the surface magnetic field and the LSD Stokes $V$ profiles makes it unlikely that significant departures from a magnetic dipole can explain the magnetospheric asymmetry. The only other stars that exhibit this kind of highly asymmetric emission are also close binary systems, suggesting that dynamical influences may play a role in shaping the circumstellar plasma accumulation surface. 

Analysis of the {\em K2} light curve of $\rho$\,Oph\,C revealed $\rho$\,Oph\,Ab's previously unreported light curve hidden within. The light curve is consistent with expectations for an eclipsing rigidly rotating magnetosphere, which we were able to approximately reproduce using a CBO-modified RRM model. The photometric eclipses align with magnetic nulls and H$\alpha$ eclipses, as expected. The light curve's highly asymmetrical eclipse depths are consistent with the asymmetry seen in H$\alpha$ emission. While complete modeling should attempt to simultaneously reproduce the light curve, H$\alpha$, and \bz~(along with the radio emission), the excellent agreement between the observed and modeled morphology of the primary eclipse demonstrates that $\rho$\,Oph\,Ab's light curve is almost certainly magnetospheric in origin. However, the necessity of introducing an {\em ad-hoc} phase shift requires further examination.

Since $\rho$\,Oph\,A is a close, bright system, its $\sim$1~AU orbit and expected angular separation of several mas is easily resolved by modern interferometric instrumentation such as VLTI/GRAVITY or CHARA/MIRC-X. These data have been acquired and have been analyzed by \cite{2025arXiv250510038K}. The astrometric orbital data provide much more precise constraints on the orbital inclination, enabling a better test of the hypothesis that the spin and orbital axes are in fact aligned. 

\begin{acknowledgements}
This work is based on observations obtained at the Canada-France-Hawaii Telescope (CFHT) which is operated by the National Research Council of Canada, the Institut National des Sciences de l'Univers (INSU) of the Centre National de la Recherche Scientifique (CNRS) of France, and the University of Hawaii, and at the Observatoire du Pic du Midi (France), operated by the INSU; and based on observations obtained at the Dominion Astrophysical Observatory, National Research Council Herzberg Astronomy and Astrophysics Research Centre, National Research Council of Canada. Based on observations made with ESO Telescopes at the La Silla and Paranal Observatories under programme IDs 60.A-9800(D), 71.C-0068(A), 073.C-0138(A), 082.C-0005(B), and 187.D-0917(D). This work has made use of the VALD database, operated at Uppsala University, the Institute of Astronomy RAS in Moscow, and the University of Vienna. This research has made use of the SIMBAD database, operated at CDS, Strasbourg, France. MES acknowledges the financial support provided by the Annie Jump Cannon Fellowship, supported by the University of Delaware and endowed by the Mount Cuba Astronomical Observatory. AuD acknowledges NASA ATP grant number 80NSSC22K0628 and support by NASA through Chandra Award number TM1-22001B and GO2-23003X issued by the Chandra X-ray Observatory 27 Center, which is operated by the Smithsonian Astrophysical Observatory for and on behalf of NASA under contract NAS8-03060. BD acknowledges support from the Bartol Research Institute. GAW acknowledges Discovery Grant support from the Natural Sciences and Engineering Research Council (NSERC) of Canada.
\end{acknowledgements}

\section*{Data Availability Statement}

Reduced ESPaDOnS spectra are available at the CFHT archive maintained by the CADC\footnote{\url{https://www.cadc-ccda.hia-iha.nrc-cnrc.gc.ca/en/}}, where they can be found via standard stellar designations. ESPaDOnS and Narval data can also be obtained at the PolarBase archive\footnote{\url{http://polarbase.irap.omp.eu/}}. {\em Kepler-2} data are available at the MAST archive\footnote{\url{https://mast.stsci.edu/portal/Mashup/Clients/Mast/Portal.html}}. UVES data are available at the ESO archive\footnote{\url{https://archive.eso.org/cms/data-portal.html}}. Additional spectra (HARPS-N, CAOS, DAO) are available from the authors on request.

\bibliography{bib_dat}{}

\appendix

\section{tables of measurments}\label{appendix:bz}

\begin{table}
\centering
\caption[]{RV measurements. The number of spectra in each 10-day bin are given by $N$. RV uncertainties are estimated at 10 \kms.}
\label{table:rv}
\begin{tabular}{l l l r r r}
\hline\hline
Calendar & HJD          & Instrument & $N$ & RV$_{\rm Aa}$  &  RV$_{\rm Ab}$ \\
Date     &  $- 2450000$ &            &     & (${\rm km/s}$) & (${\rm km/s}$) \\               \\
\hline
21/05/05 & 3511 & ESPaDOnS &  2 &  53 & -71 \\
04/09/05 & 3618 & UVES &  3 & -44 &   0 \\
10/07/14 & 6849 & Narval &  8 &  56 & -84 \\
29/05/18 & 8267 & CAOS &  4 &  45 & -48 \\
21/06/18 & 8290 & CAOS &  4 & -50 &  56 \\
27/06/18 & 8296 & HARPS-N &  4 & -55 &  49 \\
06/07/18 & 8305 & CAOS &  2 & -63 &  22 \\
16/07/18 & 8315 & CAOS &  2 & -50 &   0 \\
30/07/18 & 8330 & CAOS &  4 &  -2 & -60 \\
02/08/18 & 8333 & CAOS &  4 &   1 & -66 \\
22/05/21 & 9356 & ESPaDOnS & 11 & -49 &  44 \\
22/06/21 & 9388 & DAO &  4 &  32 & -84 \\
20/07/21 & 9415 & DAO &  2 &  10 & -37 \\
\hline\hline
\end{tabular}
\end{table}

\begin{table*}[h]
\centering
\caption[]{\bz~measurements. DF indicates detection flag (see text). Superscripts $c$ next to the instrument indicate co-added observations.}
\label{table:bz}
\begin{tabular}{l r l r r r r}
\hline\hline
Calendar Date & HJD $- 2450000$        & Instrument & \bz~(G)   & DF$_V$ & \nz~(G) & DF$_N$ \\
\hline
\\
10/07/2014 & 6849.39507 &  Narval$^c$ & $  -25 \pm   203$ &  ND & $  120 \pm   203$ &  ND \\
18/05/2021 & 9352.97194 &  ESPaDOnS & $  -19 \pm   156$ &  ND & $  113 \pm   156$ &  ND \\
19/05/2021 & 9353.89960 &  ESPaDOnS & $ 1001 \pm   118$ &  DD & $ -131 \pm   118$ &  ND \\
19/05/2021 & 9353.92533 &  ESPaDOnS & $ 1171 \pm   107$ &  DD & $ -202 \pm   107$ &  ND \\
20/05/2021 & 9354.94715 &  ESPaDOnS & $ -288 \pm   120$ &  DD & $   45 \pm   120$ &  ND \\
21/05/2021 & 9355.94507 &  ESPaDOnS & $  -88 \pm   677$ &  ND & $  492 \pm   677$ &  ND \\
21/05/2021 & 9355.95738 &  ESPaDOnS & $ -230 \pm   166$ &  ND & $ -184 \pm   166$ &  ND \\
23/05/2021 & 9357.95048 &  ESPaDOnS & $ -723 \pm   121$ &  DD & $   19 \pm   120$ &  ND \\
24/05/2021 & 9358.91434 &  ESPaDOnS & $ -177 \pm   149$ &  ND & $   -7 \pm   149$ &  ND \\
26/05/2021 & 9360.86126 &  ESPaDOnS & $  214 \pm   106$ &  DD & $    8 \pm   106$ &  ND \\
27/05/2021 & 9361.83993 &  ESPaDOnS & $ -557 \pm   126$ &  ND & $  197 \pm   126$ &  ND \\
27/05/2021 & 9362.04128 &  ESPaDOnS & $  562 \pm   116$ &  DD & $  178 \pm   116$ &  ND \\
28/05/2021 & 9363.04225 &  ESPaDOnS & $  651 \pm    92$ &  DD & $   23 \pm    92$ &  ND \\
\hline\hline
\end{tabular}
\end{table*}

\end{document}